\documentclass[12pt]{article}

\pdfoutput=1

\usepackage[top=80pt,bottom=85pt,left=67pt,right=67pt]{geometry}
\usepackage{amssymb,amsmath,xypic}
\usepackage{graphicx,color,subfigure}
\usepackage{float}
\usepackage{cite}
\usepackage[debug,pageanchor=false]{hyperref}
\definecolor{link}{rgb}{.8,.15,.1}
\hypersetup{colorlinks=true,linkcolor=link,citecolor=link,urlcolor=link,linktocpage}

\makeatletter
\@addtoreset{equation}{section}
\makeatother

\newcommand{\beq}{\begin{equation}}
\newcommand{\eeq}{\end{equation}}
\newcommand{\bea}{\begin{eqnarray}}
\newcommand{\eea}{\end{eqnarray}}
\newcommand{\nn}{\nonumber}

\begin{document}

\begin{titlepage}

\begin{center}

\vskip .5in 
\noindent

{\Large \bf AdS$_3$ solutions with $\mathcal{N}=(3,0)$ from S$^3\times$S$^3$ fibrations }

\bigskip\medskip

Andrea Legramandi$^{a,}$\footnote{a.legramandi@campus.unimib.it}, Niall T. Macpherson$^{b,}$\footnote{ntmacpher@gmail.com}\\

\bigskip\medskip
{\small

$a$: Dipartimento di Fisica, Universit\`a di Milano--Bicocca, \\ Piazza della Scienza 3, I-20126 Milano, Italy \\ and \\ INFN, sezione di Milano--Bicocca\\
\vskip 3mm
 $b$: International Institute of Physics, Universidade Federal do Rio Grande do Norte,
Campus Universitario - Lagoa Nova, Natal, RN, 59078-970, Brazil}

\vskip .9cm 
     	{\bf Abstract }

\vskip .1in
\end{center}

\noindent
We study warped AdS$_3$ solutions in massive IIA supergravity preserving $\mathcal{N}=(3,0)$ and $\mathcal{N}=(1,0)$ supersymmetry. We consider solutions whose internal spaces decompose as an S$^3\times$S$^3$ fibration and a interval over which the rest of the solution is foliated. We present necessary and sufficient conditions for these solutions to exist, in terms of systems of ordinary differential equations  and find several new analytic and numerical examples with internal spaces bounded between D-brane and O-plane behaviors.  

\noindent

\vfill
\eject

\end{titlepage}

\tableofcontents

\section{Introduction}
The AdS$_3/$CFT$_2$ correspondence is the arena in which we can best hope to test the holographic paradigm. In part this is due to the relative tractability of CFTs in two dimensions, moreover there has also been significantly more progress made towards quantising string theory in AdS$_3$ backgrounds (see the seminal works \cite{Maldacena:2000hw,Maldacena:2000kv,Maldacena:2001km}) than in higher dimensional AdS cases. This provides powerful tools to probe the AdS-CFT correspondence, even beyond the strict large N limit of CFTs and classical limit of supergravity (see the recent works \cite{Gaberdiel:2018rqv,Eberhardt:2019ywk,Dei:2019osr,Eberhardt:2019qcl} and references therein).

Superconformal field theories in two dimensions have a rich structure of possible superconformal algebras associated to them, this is contrary to higher dimensional examples where the number of preserved supercharges uniquely fixes the associated algebra. The classification and construction of holographic duals realising this vast array of algebras is certainly an interesting problem which is still largely unknown (however \cite{Martelli:2003ki,Kim:2005ez,Gauntlett:2006af,Gauntlett:2006ns,Figueras:2007cn,	Donos:2008hd,DHoker:2008lup,Colgain:2010wb,Estes:2012vm,Bachas:2013vza} for early classification results). Recently, more attention has been given to populating the space of supergravity solutions with various superconformal algebras. Most efforts have focused on type IIB  with partial results for solutions with small $\mathcal{N}=(4,0)$ in \cite{Couzens:2017way,Couzens:2019wls}, large $\mathcal{N}=(4,0)$ in \cite{Macpherson:2018mif,Dibitetto:2017klx} and $\mathcal{N}=(2,0)$ in \cite{Couzens:2017nnr,Couzens:2018wnk,Couzens:2019iog,	Couzens:2019mkh} and $\mathcal{N}=(1,0)$ in \cite{Passias:2019rga}. Exceptions to this trend include a classification of purely NS $\mathcal{N}=(2,2)$ solutions in \cite{Eberhardt:2017uup}, large $\mathcal{N}=(4,0)$ in M-theory \cite{Kelekci:2016uqv} and massive IIA \cite{Macpherson:2018mif}, and small $\mathcal{N}=(4,0)$ in massive IIA \cite{Lozano:2019emq,Lozano:2019jza,Lozano:2019zvg,Lozano:2019ywa}. These examples, while certainly of great merit, still only cover a small subset of possible superconfromal algebras - see \cite{Beck:2017wpm} for a complete list that may be embedded into ten- and eleven-dimensional supergravity \footnote{Interestingly \cite{Deger:2019tem} find $\mathcal{N}=(8,0)$ solutions in 3d gauged supergravity that do not appear to be able to be embedded in 10 or 11 dimensions - so this may not be the full story.}. Rather less ``vanilla'' options, were presented in \cite{Dibitetto:2018ftj} where solutions with F(4) and G(3) supergroups were constructed. In this paper we aim to expand on this story and construct $\mathcal{N}=(3,0)$ solutions in massive IIA preserving the supergroup OSP(3$|$2) ( see \cite{Figueras:2007cn} for an earlier study in the context of M-theory).

Our strategy for constructing solutions preserving $\mathcal{N}=(3,0)$ supersymmetry will be to construct spinors and bosonic fields which manifestly realise the bosonic subgroup of OSP(3$|$2), namely SL$(2,\mathbb{R})\times$SO(3). Specifically, we will demand that the bosonic fields are SL$(2,\mathbb{R})\times$SO(3) singlets while the spinors transform in the (\textbf{2},\textbf{3}) representation.
The SL$(2,\mathbb{R})$ symmetry is ensured with a (warped) AdS$_3$ factor in the metric and by decomposing the ten-dimensional Majorana--Weyl Killing spinors as a product of Killing spinors on AdS$_3$ and an internal seven-manifold. Realising the SO(3) R-symmetry, SO(3)$_R$, is a little more tricky, as there is no symmetric space (i.e. S$^n$,T$^n$,H$^n$,...) whose Killing spinors transform as a triplet under SO(3)$_R$. Taking inspiration from \cite{DeLuca:2018buk}, we get around this issue by instead using spinors on the internal space that realise an SO(4) R-symmetry, which can be achieved with a product of 2 and/or 3-sphere (we choose 3-spheres building on the work of \cite{Macpherson:2018mif}), and then explicitly breaking SO(4) to SO(3) with the fluxes and a S$^3\times$S$^3$ fibration. An advantage of this approach is that in addition to necessary conditions for solutions with OSP(3$|$2), we also find necessary conditions for a class of $\mathcal{N}=(1,0)$ solutions with OSP(1$|$2) superconformal algebra, for which SO(3) becomes a flavour symmetry. \\
~\\
The layout of the paper is as follows: In section \ref{sec:one} we review the construction of SO(4) spinors on a foliations of AdS$_3\times$S$^3\times$S$^3$ over an interval and explicitly spell out how we break SO(4) to SO(3) with the fluxes. We also explain why this breaking of symmetry leads to both $\mathcal{N}=(3,0)$ and $\mathcal{N}=(1,0)$ solutions. In section \eqref{sec:SUSYconds} we then use the necessary geometric conditions for  AdS$_3$ solutions in massive IIA to preserve supersymmetry presented in \cite{Dibitetto:2018ftj}. We arrive at systems of ordinary differential equations (ODEs) whose solutions define supergravity backgrounds preserving $\mathcal{N}=(3,0)$ and $\mathcal{N}=(1,0)$ supersymmetry. Finally in sections \ref{sec:exneq3} and \ref{sec:exneq1}, we present several analytic and numerical solutions to these ODEs that bound the internal seven-manifold between various D-brane and O-plane behaviors. As such, they constitute good candidates for holographic duals to two-dimensional SCFTs.

\section{Realising $\mathcal{N}=(3,0)$ from AdS$_3\times$S$^3\times$S$^3\times \mathbb{R}$ with a fibration}\label{sec:one}
Our main goal is to construct $\mathcal{N}=(3,0)$ AdS$_3$ backgrounds in massive IIA supergravity. As such we consider solutions with bosonic fields that may be decomposed as
\begin{align}\label{eq:bosonicfields}
ds^2&= e^{2A} ds^2(\text{AdS}_3)+ ds^2(\text{M}_7),\nn\\[2mm]
F&=  f+ e^{3A}\text{vol}(\text{AdS}_3)\wedge\star_7 \tilde{\lambda}(f),
\end{align}
where $F$ is the RR polyform\footnote{I.e. $F= F_0+ F_2+ F_4+ F_6 + F_8 + F_{10}$ where $F_6=-\star_{10} F_4,~F_8=\star_{10} F_2,~F_10=-\star_{10} F_0$.}, $f$ is its magnetic components with legs and functional support on $\text{M}_7$ only, and the function $\tilde{\lambda}$ acts on a n-form as $\tilde{\lambda}(X_n)= (-1)^{\frac{n(n-1)}{2}} X_n$. The AdS warp factor $e^{2A}$ and dilaton $\Phi$ have support on M$_7$ only and the NS 3-form $H$, like $f$, is purely magnetic\footnote{This is a requirement for non trivial Romans mass F$_0$ \cite{Dibitetto:2018ftj}.}. The ten-dimensional Majorana--Weyl Killing spinors of such a background decompose as
\beq\label{eq10dspinors}
\epsilon_1=\sum_{i=1}^3 \zeta^i \otimes v_+\otimes \chi^i_1,~~~~\epsilon_2= \sum_{i=1}^3\zeta^i \otimes v_{-}\otimes \chi^i_2 
\eeq
with  $\zeta^I$ independent Majorana Killing spinors on AdS$_3$ which obey
\beq
\nabla_{M}\zeta= \frac{\mu}{2}\gamma^{(2)}_M \zeta,~~~~\mu= \pm 1,
\eeq
and are charge under the SL(2)$_{\pm}$ factor of SO(1,2)$\sim$SL(2)$_+\times$SL(2)$_-$. $\chi^I_{1,2}$ are also independent Majorana spinors on M$_7$ and the remaining objects $v_{\pm}$ are auxiliary 2-vectors that are always required when decomposing an even dimensional spinor in terms of two odd dimensional ones - they guarantee $\epsilon_{1,2}$ are a representation of Cliff(1,9) and take care of the ten-dimensional chirality\footnote{We refer to \cite{Macpherson:2018mif} for more details about the Clifford algebra decomposition.}. \\
~\\
Solutions with $\mathcal{N}$=(3,0) superconformal symmetry should realise the supergroup OSP(3$|$2); to ensure this it is sufficient to manifestly realise its bosonic subgroup SL(2)$\times$SO(3) with the bosonic fields \eqref{eq:bosonicfields} and Killing spinors \eqref{eq10dspinors}. Specifically, the bosonic fields should be  SL(2)$\times$SO(3) singlets, while the spinors transform in the (\textbf{2},\textbf{3}). The SL(2) factor is realised by the AdS$_3$ Killing spinors, and it is clearly respected by the bosonic fields \eqref{eq:bosonicfields}. The SO(3) factor is an R-symmetry (SO(3)$_\text{R}$) that must be implemented in the internal-space geometry: this restricts the possible local forms M$_7$ and $\chi^{i}_{1,2}$ can take. In particular, $\chi^{i}_{1,2}$ must be an SO(3) triplet, so it is possible to parameterise the internal spinors such that their Lie derivative along the SO(3) Killing vectors K$_i$ obey
\beq
\mathcal{L}_{K_i}\chi^j_{1,2}=\epsilon_{ijk}\chi^k_{1,2}.
\eeq
The easiest way to realise an SO(3) isometry of the bosonic fields is with a round two-sphere, however the Killing spinors on S$^2$ transform in  the \textbf{2} of SU(2), which is not what we want - the situation does not improve with the 3-sphere whose two independent Killing spinors transform in the $(\textbf{2},0)$ and $(0,\textbf{2})$ of SU(2)$\times$SU(2).  Instead, one can realise SO(3)$_{\text{R}}$ by first constructing spinors that realises an SO(4) R-symmetry and then breaking this down to SO(3)$_{\text{R}}$ with the bosonic fields.
 To realise SO(4), one needs to consider spinors on one of S$^2\times $S$^2,~$S$^2\times $S$^3$ or S$^3\times S^3$. The general form of SO(4) spinors on S$^3\times$S$^3\times \mathbb{R}$ was already given in \cite{Macpherson:2018mif}, so this shall be our starting point. We review their construction in the next section.\\

\subsection{Constructing SO(4) spinors on AdS$_3\times$S$^3\times $S$^3\times\mathbb{R}$}
In \cite{Macpherson:2018mif} the general form of SO(4) spinors on AdS$_3\times$S$^3\times$S$^3\times\mathbb{R}$ was derived and then used to  find the local form of all such type II solutions. The purpose of this section is to review this first point.

We begin by imposing some additional structure on \eqref{eq:bosonicfields} and the remaining bosonic fields. We decompose the internal manifold M$_7$ as 
\beq\label{eq:originalM7}
ds^2(\text{M}_7)= e^{2C_1} ds^2(S^3_1)+ e^{2C_2} ds^2(S^3_2)+ e^{2k}d\rho^2,~~~~ 
\eeq
and constrain the fluxes $f,H$ to depend on the 3-sphere directions only through their respective volume forms vol(S$^3_{1,2}$), which span all possible SO(4) invariant forms on S$^3\times$S$^3$.  Additionally $f,H,\Phi,A,C_{1,2},k$  have function support in the interval $\rho$ only. Such solutions realises an SO(4)$\times$SO(4) isometry and are consistent with (at least) an SO(4) R-symmetry. General spinors charged under SO(4) on this background were constructed in \cite{Macpherson:2018mif} by taking products of SU(2) doublets on the two 3-sphere, they take the form
\beq\label{eq:SO(4)}
\chi^{\hat{I}}_1=e^{\frac{A}{2}}\left(\begin{array}{c}\sin(\alpha_1+\alpha_2)\\i\cos(\alpha_1+\alpha_2)\end{array}\right)\otimes \eta^{\hat{I}},~~~~\chi^{\hat{I}}_2=e^{\frac{A}{2}}\left(\begin{array}{c}\sin(\alpha_1-\alpha_2)\\i\cos(\alpha_1-\alpha_2)\end{array}\right)\otimes \eta^{\hat{I}},
\eeq
where $I=1,..4$ and $\alpha_{1,2}$ are $\rho$ dependent phases, to be fixed by the necessary conditions for supersymmetry and Bianchi identities of the RR fluxes. The spinor $\eta^{\hat{I}}$ is defined as
\beq
\eta^{I}=  (\mathcal{M}_{I})_{ab} \xi_1^a\otimes \xi_2^b,~~~~\mathcal{M}_{I}= (\sigma_2 \sigma_1,\sigma_2 \sigma_2,\sigma_2 \sigma_3,-i\sigma_2)_{I},
\eeq
for $a=1,2$. Here $\xi_{1,2}^a$ are SU(2) doublets on S$^{3}_{1,2}$ (see \cite{Macpherson:2017mvu} for details) defined in terms of the 3-sphere Killing spinors as
\beq
\xi_{1,2}^a=\left( \begin{array}{c}\xi_{1,2}\\\xi^c_{1,2}\end{array}\right)^a~~~\text{where}~~~\nabla_{\beta_{1,2}}\xi_{1,2}=\nu\frac{i}{2}\gamma_{\beta_{1,2}}\xi_{1,2},~~~~ \nu=\pm 1,
\eeq 
where $\beta_{1,2}=1,..3$ are coordinates on the unit 3-spheres, and $\xi^c$ denotes the Majorana conjugate of $\xi$. The parameter $\nu=\pm 1$ determines which factor of SO(4)$\sim$SU(2)$_+\times$SU(2)$_-$ the Killing spinors are charged under.

The SO(4) R-symmetry is embedded in SO(4)$\times$SO(4) as follows: the 3-sphere S$^3_{1,2}$ admits a global SO(4)$_{1,2}\sim$SU(2)$_{1,2+}\times$SU(2)$_{1,2,-}$ isometry. Let us assume for simplicity that $\nu=1$\footnote{analogous statements hold for $\nu=-1$, but the sign of $\nu$ only holds any physical significance when both values are allowed by a solution. In this case the R-symmetry is enhanced to SO(4)$\times$SO(4) } so that $\xi^{\alpha}_{1,2}$ is charged under SU(2)$_{1,2+}$, the SO(4) R-symmetry is then SO(4)$_{\text{R}}$= SO(3)$_{\text{D}}\times$SO(3)$_{\text{AD}}$ where SO(3)$_{(\text{A})\text{D}}$ is the (anti-)diagonal subgroup of SU(2)$_{1+}\times$SU(2)$_{2+}$ with Killing vectors
\beq
K^{\text{D}}_{i}= K_i^1+ K^2_i,~~~~ K^{\text{AD}}_i= K_i^1-K_i^2
\eeq
where $K^{1,2}_{i}$ are the SU(2) Killing vectors on $S^3_{1,2}$ which in general, as 1-forms, obey
\beq
dK^{1,2}_i=\frac{\nu}{2} \epsilon_{ijk}K^{1,2}_{j}\wedge K^{1,2}_{k}.
\eeq
The SO(4) spinors  $\chi^I_{1,2}$ both transform under the spinorial Lie derivative as
\beq\label{eq:SO4action}
\mathcal{L}_{K^{\text{D}}_i}\chi^I_{1,2}= \nu \left(\begin{array}{c|c}\epsilon_{ijk}& \underline{0}\\\hline \underline{0}^T &0\end{array}\right)^I_{~J}\chi^J_{1,2},~~~~ \mathcal{L}_{K^{\text{AD}}_i}\chi^I_{1,2}= \nu \left(\begin{array}{c|c}0_{3\times 3}& \underline{c_i^T}\\\hline -\underline{c_i} &0\end{array}\right)^I_{~J}\chi^J_{1,2},
\eeq
where $\underline{c_1}=(1,0,0),~\underline{c_2}=(0,1,0),~\underline{c_3}=(0,0,1)$. This makes it clear that if we decompose $\chi^I_{1,2}=(\chi^i_{1,2},\chi^4_{1,2})$ for $i=1,2,3$, then $\chi^i_{1,2}$ is an SO(3)$_{\text{D}}$ triplet while $\chi^4_{1,2}$ is an SO(3)$_{\text{D}}$ singlet. We can thus break supersymmetry to $\mathcal{N}=(3,0)$, by breaking the SO(3)$_{\text{AD}}$ symmetry with the bosonic fields. 

Let us stress that it was proved in \cite{Macpherson:2018mif} (see Appendix B therein) that when the bosonic fields are singlets under the SO(4) R-symmetry, it is sufficient to solve the necessary conditions for supersymmetry that follow from a single component of $\chi^I_{1,2}$, as the others automatically follow through the action of SO(4) through \eqref{eq:SO4action}. If we break the SO(3)$_{\text{AD}}$ symmetry of SO(4)=SO(3)$_{\text{D}}\times $SO(3)$_{\text{AD}}$, leaving SO(3)$_{\text{D}}$ intact, things are a little different. Solving the supersymmetry conditions that follow from any of $(\chi^1_{1,2},\chi^2_{1,2},\chi^3_{1,2})$ will give solutions preserving $\mathcal{N}=(3,0)$ supersymmetry. If we instead solve the conditions that follow from $\chi_{1,2}^4$, then the solution will preserve just $\mathcal{N}=(1,0)$ and SO(3)$_{D}$ becomes a flavour symmetry - this is essentially because \eqref{eq:SO4action} only mixes $\chi^4_{1,2}$ with $\chi^i_{1,2}$ through  SO(3)$_{\text{AD}}$, which we choose to break.\\
~\\
In the next section we will spell out precisely how we will break SO(3)$_{\text{AD}}$.

\subsection{Partially breaking $\mathcal{N}=(4,0)$ with an S$^3\times$S$^3$ fibration}
In this section we will partially break the $\mathcal{N}=(4,0)$ supersymmetry ansatz of the previous section with the bosonic fields. The easiest way to do this is with an orbifold, see for instance \cite{Eberhardt:2018sce}, where orbifolds of AdS$_3\times$S$^3\times$S$^3\times$S$^1$ that preserve $\mathcal{N}=(3,3)$ and $\mathcal{N}=(1,1)$ are considered. However this only break supersymmetry globally, and in particular the local form of a solution and its orbifold are the same. Here we would like to break supersymmetry in a more dramatic fashion - there are two options available to us: break with the fluxes or break with the metric, here we will do both.\\
~\\
To partially break supersymmetry with the metric we can fibre S$^3_1$ over S$^3_2$ in such a way that we manifestly break SO(3)$_{\text{AD}}$, which leads us to modify \eqref{eq:originalM7} as
\beq\label{eq:intfibred}
ds^2(\text{M}_7)= \frac{e^{2C_1}}{4}\big(K^1_i+ \lambda K^2_i\big)^2+ \frac{e^{2C_2}}{4} \big(K^2_i\big)^2+ e^{2k}d\rho^2, 
\eeq
with $\lambda=\lambda(\rho)$ an arbitary function. Let us now also fix the Killing spinor parameters $\mu=\nu =1$, so that M$_7$ preserves a flavour SO(3)$_{1,-}\times$SO(3)$_{2,-}$ and the diagonal SO(3) subgroup of SO(3)$_{1,+}\times$SO(3)$_{2,+}$, which may be an R-symmetry or a flavour symmetry, depending on our spinor ansatz.\\
~\\
To break the SO(4) R-symmetry with the fluxes, we simply need to allow them to depend on the forms on S$^3_1\times$S$^3_2$ that are invariant under SO(3)$_{\text{D}}$ but not SO(3)$_{\text{AD}}$ (in addition to the  SO(4) invariant forms). A basis of all SO(3)$_{\text{D}}$ invariant forms is given by
\begin{align}\label{eq:invforms}
\omega_2&=K^F_i\wedge K^2_i,\\[2mm]
\omega^1_3&= \frac{1}{8}K^F_1\wedge K^F_2\wedge K^F_3,~~~\omega^2_3= \frac{1}{8}K^2_1\wedge K^2_2\wedge K^2_3,\\[2mm]
\omega^3_3&= \frac{1}{16}\epsilon_{ijk} K^1_i\wedge K^F_j\wedge K^F_k,~~~\omega^4_3= \frac{1}{16}\epsilon_{ijk} K^F_i\wedge K^2_j\wedge K^2_k,\\[2mm]
\omega_4&= \frac{1}{16} dK^1_i\wedge dK^2_i=-\frac{1}{2}\omega_2\wedge \omega_2,~~~\omega_6=\omega^1_3\wedge \omega^2_3=-\frac{1}{6}\omega_2\wedge\omega_2\wedge\omega_2=-\frac{1}{3}\omega^3_3\wedge \omega^4_3,
\end{align}
where to lighten the notation we have defined
\beq
K^F_i=K^1_i+ \lambda K^2_i,
\eeq
and unless the converse is stated above, the invariant forms vanish when wedged with each other. They also form a closed set under exterior differentiation, namely
\begin{align}
d\omega_2&= 6\lambda(1+ \lambda)\omega^1_2+ 2\omega^3_3-2(1+ 2\lambda)\omega^4_3,\nonumber\\[2mm]
d\omega^1_3&= 2(1+ \lambda) \lambda \omega_4+ \partial_{\rho}\lambda d\rho\wedge \omega^3_3,\nn\\[2mm]
d\omega^3_3&=2(1+2\lambda)\omega_4+ 2\partial_{\rho}\lambda d\rho\wedge \omega^4_3,\nn\\[2mm]
d\omega^4_3&=2\omega_4+ 3\partial_{\rho}\lambda d\rho\wedge\omega^2_3,
\end{align}
with $d$ of all else yielding zero.
Notice that there exists two 3-forms which are closed but not exact
\beq
\text{vol}(S^3_1)=\omega^1_3-\lambda^3\omega^2_3+\lambda^2\omega^4_3-\lambda \omega^3_3,~~~\text{vol}(S^3_2)=\omega^2_3, 
\eeq
 which as the notation suggests, give the volume forms of each un-fibered 3-spheres. Given the forms at our disposal we may expand the NS three-form as
\beq
H= q_1 \omega^1_3+q_2 \omega^2_3+p \omega^3_3 +q_3 \omega^4_3+ q_4 d\rho\wedge \omega_2,
\eeq
where $(q_i,p)$ are each arbitrary functions of $\rho$. Solving the Bianchi identity $d H = 0$ away from potential localised sources then imposes
\begin{align}
q_1&= c_1,~~~ q_2= c_2+ \lambda(3 p(1+ \lambda)+ c_1 \lambda(3+ 2\lambda)),\\[2mm]
q_3&= -c_1 \lambda(1+ \lambda)-p(1+ 2\lambda),~~~ q_4=\partial_{\rho}\frac{p+ c_1 \lambda}{2}
\end{align} 
where $c_{1,2}$ are constants. This allows to write the general local form of the NS three-form as
\beq\label{eq:NS3-form}
H= d(\frac{1}{2}(p+c_1 \lambda)\omega_2)+ c_1 \text{vol}(S^3_1)+ c_2 \text{vol}(S^3_2).
\eeq
Similar expressions exist for the RR forms, we can expand their magnetic components in terms of seven functions of $\rho$, $u_1,...,u_7$ as
\begin{align}\label{eq:RRansatz}
f_0&=F_0,~~~~f_2= u_1 \omega_2,~~~~f_6= u_7 \omega^3_1\wedge \omega^3_2,\nn\\[2mm]
f_4&=  d\rho \wedge(u_2 \omega^1_3+ u_3\omega^2_3+u_4\omega^3_3+u_5\omega^4_3)+ u_6\omega_4.
\end{align}
The electric components of the RR fluxes are defined in terms of \eqref{eq:RRansatz} as in \eqref{eq:bosonicfields}, which requires one to take the hodge dual on M$_7$ - we quote the necessary identities to achieve this in appendix \ref{hodge}.
 
Away from the loci of possible delta-function sources, the fluxes should obey the Bianchi identities $dF_n=H\wedge F_{n-2}$, but as we are interested in solutions preserving at least $\mathcal{N}=(1,0)$ supersymmetry, the electric contributions will be implied leaving only the magnetic contributions to solve for. Along with requiring that $F_0$ is constant\footnote{In general this is need be true piece-wise, with discontinuities signaling  D8 sources.} they imply the following constraints,
\begin{align}\label{eq:bis}
u_1&=\frac{F_0}{2}(p+ c_1 \lambda),~~~c_1F_0=c_2F_0=0,\\[2mm]
u_5&=\frac{1}{2}\partial_{\rho}(\frac{F_0}{4}p^2+u_6)-u_2\lambda(1+\lambda)-u_4(1+2 \lambda),\nn\\[2mm]
4c_1u_3&=4u_2(c_2+3c_1 \lambda^2+2 c_1 \lambda^2)+3\big(4c_1 u_4\lambda(1+\lambda)+F_0 p^2 \partial_{\rho}p+2 c_1 u_6 \partial_{\rho}\lambda\big)-4 u_7'+ 6\partial_{\rho}(p u_6),\nn
\end{align}
which line by line follow from the Bianchi identities for $f_2,f_4,f_6$, simplified with the proceeding conditions.
\\
~\\
In the next section we derive two sets of ODEs that imply supersymmetry is preserved. When these are solved along with the Bianchi identities \eqref{eq:bis}, they give rise to solutions preserving either $\mathcal{N}=(3,0)$ or $\mathcal{N}=(1,0)$ supersymmetry.

\section{Necessary conditions for AdS$_3$ solutions to exist}\label{sec:SUSYconds}
Recently \cite{Dibitetto:2018ftj} provided necessary and sufficient conditions for $\mathcal{N}=(1,0)$ supersymmetry to be preserved by a solution in massive IIA. The fundamental object in terms of which these conditions are formulated is the seven-dimensional bi-spinor defined in terms of two Majorana spinors $\chi_{1,2}$ as
\beq\label{eq:bispinor}
\chi_1\otimes\chi^{\dag}_2 = \frac{1}{8}\sum_{n=0}^7\frac{1}{n!} \chi_2^{\dag}\gamma_{a_n...a_1} \chi_1 e^{a_1}\wedge .... \wedge e^{a_n},
\eeq
where $\gamma_{a}$ is basis of seven-dimensional flat-space gamma matrices and $e^a$ is the veilbein on M$_7$. In general the bi-spinor decomposes in terms of it's even/odd parts labeled by $\pm$ as
\beq
\chi_1\otimes\chi^{\dag}_2=\Psi_++ i \Psi_-,
\eeq
for $\Psi_{\pm}$ both real. Supersymmetry is then ensured in IIA when the following conditions hold
\begin{align}\label{eq:susycond7d}
&(d-H\wedge)(e^{A-\Phi}\Psi_{-})=0,\nn\\[2mm]
&(d-H\wedge)(e^{2A-\Phi}\Psi_{+})-2 \mu e^{A-\Phi}\Psi_{-}= \frac{e^{3A}}{8}\star_7\tilde{\lambda}(f),\nn\\[2mm]
&e^{-\Phi}(f,\Psi_{+})-\frac{\mu}{2} \text{Vol}_7=0,
\end{align}
with $(.~,~.)$  the Mukai pairing in seven dimensions defined as $
(X,Y)=  \bigg(\tilde\lambda(X)\wedge Y\bigg)_7$. Specifically these are the conditions when one assumes that
\beq
|\chi_1|^2=|\chi_2|^2 = e^{A},
\eeq
which is a requirement for non-zero Romans mass. Though this need not to hold in general, we will restrict our considerations to solutions of this type.\\
~\\
In order to use this formalism we need to define $\chi_{1,2}$ on the fibred and foliated internal space \eqref{eq:intfibred}.  As argued in the final paragraph below \eqref{eq:SO4action}, this amounts to choosing which component of the SO(4) spinors \eqref{eq:SO(4)} we solve for - with any of $(\chi^1_{1,2},\chi^2_{1,2},\chi^3_{1,2})$ leading to solutions that preserve $\mathcal{N}=(3,0)$ supersymmetry and $\chi^4_{1,2}$ leading to solutions that preserve just  $\mathcal{N}=(1,0)$. As such we will take our representative $\mathcal{N}=1$ sub sectors of spinors to be $\chi_{1,2}=\chi^1_{1,2}$ in the next section where we will derive $\mathcal{N}=(3,0)$ conditions. While in section \ref{sec:neq1conds} we will take   $\chi_{1,2}=\chi^4_{1,2}$ and derive  $\mathcal{N}=(1,0)$ conditions. In both cases we need to construct the seven-dimensional bi-spinors using \eqref{eq:bispinor}, so it is helpful to know the bi-spinor relation for a round three-sphere of radius $e^{C_{1,2}}$
\beq\label{eq:3-spherebi}
\xi^a_{1,2} \otimes \xi^{b\dag}_{1,2}= \frac{1}{2}\bigg(\big(1- i e^{3C_{1,2}}\text{Vol}(\text{S}^3_{1,2})\big)\delta^{ab}+\big(\frac{1}{2}e^{C_{1,2}} K^{{1,2}}_i-\frac{i}{8}e^{2C_{1,2}}\epsilon_{ijk}K^{1,2}_j\wedge K^{{1,2}}_k\big)(\sigma^i)^{ab}\bigg)
\eeq
which is computed in \cite{Macpherson:2017mvu}. This is necessary in the derivation of $\Psi_{\pm}$ which decomposes as in terms of wedge products of bi-spinors on the two S$^3$ and the interval direction. In the case of the fibred sphere S$^3_1$ one simply needs to replace $K_i^1$ with $K_i^1+ \lambda K^2_i$. One then needs to derive ODEs on the interval spanned by $\rho$ which imply \eqref{eq:susycond7d}, under the assumption that the NS three-form is given by \eqref{eq:NS3-form} and the RR fluxes depend on the three-sphere directions exclusively through the SO(3)$_{\text{D}}$ invariant forms \eqref{eq:invforms}. This is sufficient for supersymmetry but to have a solution of IIA supergravity one needs to also impose the Bianchi identities of the RR fluxes \cite{Prins:2013wza}, then the rest of the equations of motion are implied - this amounts to solving \eqref{eq:bis}, given the flux that follows from \eqref{eq:susycond7d}.\\
~\\
In the next section we derive necessary conditions for $\mathcal{N}=(3,0)$ solutions. 

\subsection{$\mathcal{N}=(3,0)$ case}\label{eq:condtionneq3}
In this section we provide necessary a sufficient conditions for solutions preserving $\mathcal{N}=(3,0)$ supersymmetry. We begin by plugging the first components of \eqref{eq:SO(4)}, into the definition of the seven-dimensional bi-spinors \eqref{eq:bispinor}. Making use of \eqref{eq:3-spherebi}, we find that
\begin{align}
\Psi_+&=\cos\alpha_2-e^{C_1+ C_2}\sin\alpha_2  (\omega_2+4 \nu_2)-e^{2(C_1+C_2)}\cos\alpha_2(\omega_4-2\nu_4)+ e^{3(C_1+ C_2)}\omega^1_3\wedge \omega^2_3\sin\alpha_2 \nn\\[2mm]
&- e^{2k} d \rho\wedge \bigg[e^{3C_1}\cos\alpha_2\omega^1_3+e^{3C_2}\sin\alpha_2\omega^2_3+ e^{2C_1+C_2}\sin\alpha_1(\omega^3_3-2 \nu^1_3)+ e^{C_1+2C_2}\cos\alpha_1(\omega^4_3-2 \nu^2_3)\bigg],\nn
\end{align}
\begin{align}\label{eq:beq3bi}
\Psi_-&=e^k d\rho \wedge\bigg[\sin\alpha_2+ e^{C_1+C_2}\cos\alpha_2(\omega_2+4 \nu_2)-e^{2(C_1+C_2)}\sin\alpha_2(\omega_4-2\nu_4)- e^{3(C_1+C_2)}\cos\alpha_2\omega^1_3\wedge \omega^2_3\bigg]\nn\\[2mm]
& +e^{3C_1} \sin\alpha_1\omega^1_3-e^{3C_2}\cos\alpha_1\omega^2_3- e^{2C_1+ C_2}\cos\alpha_2(\omega^3_3-2 \nu^1_3)+ e^{C_1+2C_2}\sin\alpha_2(\omega^4_3-2 \nu^2_3).
\end{align}
Notice that in addition to the SO(3)$_{\text{D}}$ forms, we also have new objects appearing - they are defined as,
\begin{align}\label{eq:nus}
\nu_2&= K^F_1\wedge K^2_1,~~~\nu_4 =\frac{1}{16} dK^1_1\wedge dK^2_1,\nn\\[2mm]
\nu^1_3&= \frac{1}{16}\epsilon_{1jk} K^1_1\wedge K^F_j\wedge K^F_k,~~~\nu^2_3= \frac{1}{16}\epsilon_{1jk} K^F_1\wedge K^2_j\wedge K^2_k.
\end{align}
These are charged SO(3)$_{\text{D}}$ forms\footnote{Technically they contain both a charged and singlet contribution: One can construct matrix bi spinors $\Phi^{ij}=\chi^i_1\otimes\chi^{j\dag}_2$ which are a tensor product of SO(3) triplets and so decompose into respectively singlet, anti-symmetric and symmetric traceless irreducible representations as $\textbf{3}\otimes \textbf{3}= \textbf{1}\oplus \textbf{3}\oplus \textbf{5}$.  As it is sufficient, here we are only considering the component $\Phi^{11}$, so only the $\textbf{1}$ and (11) component of the $\textbf{5}$ appear - the forms of \eqref{eq:nus} are a linear combination of both. Though this basis does somewhat obscure the underling group theoretical structure, we choose it because it makes the following computation more simple.}, they depend on the index $1$ because we took $\chi_{1,2}=\chi^1_{1,2}$ as our explicit $\mathcal{N}=1$ sub-sector, taking $\chi_{1,2}=\chi^i_{1,2}$ leads to a dependence on $i$, but any choice of $\mathcal{N}=1$ sub-sector leads to the same necessary conditions for supersymmetry. Under exterior differentiation the charged forms behave as
\begin{align}\label{eq:chargedunderd}
d\nu_2 &= \lambda\omega^4_3-\nu^1_3+ (1-\lambda)\nu^3_3-\lambda(1+ \lambda)\omega^2_3,\nn\\[2mm]
d\nu^1_3&=2(1-\lambda)\nu_4+2 \lambda\omega_4+ \partial_{\rho}\lambda d\rho\wedge(\omega^4_3-\nu^{2}_3),\nn\\[2mm]
d\nu^2_3&=2\nu_4+ \partial_{\rho}\lambda d\rho\wedge \omega^2_3,
\end{align}
with $d\nu_4=0$. These expressions are particularly helpful when plugging \eqref{eq:beq3bi} into the supersymmetry conditions \eqref{eq:susycond7d}. To this end we also need to know the wedge products of the charged forms with the invariant 2 and 3-forms that appear in the NS 3-form, we find
\beq\label{eq:chargedunderwedge}
\nu_2\wedge \omega_2=\frac{1}{2}(\omega_4-\nu_4),~~~ \nu^1_3\wedge \omega^2_3=-\nu^2_3\wedge \omega^1_3=-\omega^1_3\wedge \omega^2_3
\eeq
with all else giving zero. Having established how all the terms appearing in \eqref{eq:NS3-form} and \eqref{eq:beq3bi} interact under exterior differentiation and wedge product, we are now ready to plug \eqref{eq:beq3bi} into \eqref{eq:susycond7d} - under the assumption that the fluxes depend only on the SO(3)$_{\text{D}}$ invariant forms- we find the following zero-form constraints
\begin{align}\label{eq:neq3zerforms}
&e^{C_1}\cos\alpha_1- e^{A} \sin\alpha_2=0,\nn\\[2mm]
&e^{C_2} \sin\alpha_1- e^{C_1}(1-\lambda)\cos\alpha_1=0,\nn\\[2mm]
&\lambda(e^{C_1} (1+ \lambda)\sin\alpha_1-e^{C_2}\cos\alpha_1)=0,\nn\\[2mm]
& c_1(e^{3C_2}\cos\alpha_1+ e^{3C_1}\lambda(\lambda^2+\lambda-1))+ c_2 e^{3C_1}\sin\alpha_1=0,
\end{align}
which are sufficient to establish that when $\lambda\neq 0$, none of  $(\cos\alpha_1,\sin\alpha_1,\cos\alpha_2,\sin\alpha_2)$ can be globally zero, as this would require also the 3-sphere or AdS warp factors to vanish- for similar reasons we must restrict to $\lambda \neq \pm 1$. One can also show that when $\lambda=0$ (i.e. when the 3-spheres are unfibred) \eqref{eq:susycond7d} implies that the fluxes only depend on the SO(4) invariant forms - thus there is an enhancement to $\mathcal{N}=(4,0)$, which is the content of \cite{Macpherson:2018mif}. As we are interested in $\mathcal{N}=(3,0)$ here,  we can safely assume $\lambda\neq -1,0,1$ and solve \eqref{eq:neq3zerforms} without loss of generality as
\beq\label{eq:neq3defs}
e^{C_1}= e^{A}\frac{\sin\alpha_2}{\cos\alpha_1},~~~ e^{C_2}= 2 e^{A}\sin\alpha_1\sin\alpha_2,~~~\lambda= \cos 2 \alpha_1,~~~c_2=-c_1=-c,
\eeq
which refines our metric ansatz to
\beq
ds^2= e^{2A}\bigg[ds^2(\text{AdS}_3)+\frac{\sin^2\alpha_2}{4 \cos^2\alpha_1}(K^i_1+ \cos 2\alpha_1 K^i_2)^2+\sin^2\alpha_1\sin^2\alpha_2(K^i_2)^2\bigg]+ e^{2k}d\rho^2,
\eeq
where $e^{2k}$ is merely a function parameterising diffeomorphism invariance on the interval.
Given the definitions \eqref{eq:neq3defs}, and after performing some tedious simplifications, we find the remaining conditions that follow form \eqref{eq:susycond7d} are: a unique definition of the function $p$  appearing in the NS 3-form
\begin{align}
p&=- 2c \cos^2\alpha_1+ 4 e^{2A}\cos\alpha_2\sin\alpha_2 \tan\alpha_1,
\end{align}
and the following system of ODEs
\begin{align}
&\partial_{\rho}(e^{A-\Phi}\tan\alpha_1 \sin\alpha_2)- F_0 e^{A+k} \tan\alpha_1=0,\label{eq:ODE-system1}\\[2mm]
&\partial_{\rho}\left(\frac{e^{5A-\Phi}\sin\alpha_1 \sin^3\alpha_2}{\cos^3\alpha_1}\right)- c e^{2A+k-\Phi}\sin\alpha_2=0,\label{eq:ODE-system2}\\[2mm]
&\partial_{\rho}(\cos 2\alpha_1)-4 e^{k-A}\cos^2\alpha_1\cot\alpha_2+2 c e^{k-3A}\frac{\cos^5\alpha_1}{\sin\alpha_1 \sin^2\alpha_2}=0,\label{eq:ODE-system3}\\[2mm]
&\partial_{\rho}(e^{2A}\tan\alpha_2)+ e^{A+k}\frac{\cos^2 2\alpha_1(1+ \cos^2\alpha_2)}{\sin^2\alpha_1\cos^2\alpha_2}- c e^{k-2A}\frac{\cos^3\alpha_1}{ \sin^3\alpha_1 \sin 2\alpha_2}=0\label{eq:ODE-system4},
\end{align}
where in the first condition we have solved the $F_0$ Bianchi identity away from localised sources by assuming it is (peice-wise) constant. We can now refine the NS 3-form as
\beq
H= db_2-\frac{c}{2} d\omega_2+ c \big(\text{vol}(\text{S}^3_1)-\text{vol}(\text{S}^3_2)\big),~~~ b_2=  e^{2A}\sin 2 \alpha_2 \tan\alpha_1\omega_2.
\eeq
In terms of this, the magnetic components of the remaining RR fluxes are then given by
\begin{align}
f_2&=  F_0 b_2,\\[2mm]
f_4&= b_2\wedge f_2-\frac{1}{2} b_2^2 F_0+\frac{4 e^{3A-\Phi}\cos\alpha_2\sin\alpha^2_2\cos 2\alpha_1}{\cos^2\alpha_1}\omega_4\nn\\[2mm]
&+e^{k-\Phi}d \rho \wedge\bigg[-\frac{c \cos^5\alpha_1 \cos\alpha_2+e^{2A}\sin\alpha_1\sin\alpha_2(3-2 \cos^2\alpha_1(2+ \cos^2\alpha_2))}{\cos^2\alpha_1\sin^3\alpha_1}\omega^1_3\nn\\[2mm]
&-8 \sin^3\alpha_1(c \cos^3\alpha_1 \cos\alpha_2+ 2 e^{2A} \sin\alpha_1 \sin^3\alpha_2)\omega^2_3\nn\\[2mm]
&+\frac{2(-c \cos^5\alpha_1 \cos\alpha_2+ e^{2A}\sin\alpha_1\sin\alpha_2(-1+ 2\cos^2\alpha_1(1+ \cos^2\alpha_2)))}{\cos^2\alpha_1\sin\alpha_1}\omega^3_3\nn\\[2mm]
&- 4 \sin\alpha_1\cos\alpha_2(c \cos^3\alpha_1-2 e^{2A}\cos\alpha_2\sin\alpha_1\sin\alpha_2)\omega^4_3\bigg],\nn\\[2mm]
f_6&= b_2\wedge f_4-\frac{1}{2}b_2^2\wedge f_2+\frac{1}{6}b_2^3F_0-\frac{8e^{3A-\Phi}\sin^2\alpha_2}{\cos^3\alpha_1}(c \cos^3\alpha_1\cos\alpha_2- e^{2A}\sin\alpha_1\sin\alpha_2)\omega^1_3\wedge \omega^2_3.\nn
\end{align}
Although these fluxes appear rather complicated, if we compare to \eqref{eq:RRansatz} to read of the functions $u_1,..u_7$, then plug these into the conditions that imply their Bianchi identities \eqref{eq:bis}, we find that only 
\beq\label{eq:remainingbi}
c F_0=0,
\eeq 
which follows from $df_2=H F_0$, is not implied by \eqref{eq:ODE-system1}-\eqref{eq:ODE-system4} - the remaining conditions $df_{n+2}=H\wedge f_n$ are implied by this when $F_0=$ constant.
Clearly \eqref{eq:remainingbi} indicates that there are two branches of solutions we should consider - we shall do so in section \ref{sec:exneq3}, though we have only found a closed form solution when both $c=F_0=0$.\\
~\\
In the next we derive necessary and sufficient conditions for $\mathcal{N}=(1,0)$ solutions.

\subsection{$\mathcal{N}=(1,0)$ case}\label{sec:neq1conds}
In this section we provide necessary and sufficient conditions for solutions preserving $\mathcal{N}=(1,0)$ supersymmetry to exist. Proceeding as before, but now with the 4th component of \eqref{eq:SO(4)} we find
\begin{align}
e^{-A}\Psi_+&=\cos\alpha_2+ e^{C_1+ C_2} \sin\alpha_2 \omega_2+ e^{2(C_1+ C_2)}\cos\alpha_2\omega_4+ e^{3(C_1+C_2)}\sin\alpha_2\omega^1_3\wedge \omega^2_3\nn\\[2mm]
&+ e^{k}d \rho \wedge\bigg[-e^{3C_1}\cos\alpha_1 \omega^1_3- e^{3C_2}\sin\alpha_1\omega^2_3+ e^{2C_1+C_2}\sin\alpha_1 \omega^3_3+ e^{C_1+2C_2}\cos\alpha_1 \omega^4_3\bigg],\nn
\end{align}
\begin{align}\label{eq:eq1bi}
e^{-A}\Psi_-&=-e^k d \rho \wedge\bigg[\sin\alpha_2- e^{C_1+ C_2}\cos\alpha_2\omega_2+ e^{2(C_1+C_2)}\sin\alpha_2\omega_4- e^{3(C_1+C_2)}\cos\alpha_2 \omega^1_3\wedge \omega^2_3\bigg]\nn\\[2mm]
&- e^{3C_1}\sin\alpha_1 \omega^1_3+ e^{3C_2}\cos\alpha_1\omega^2_3- e^{2C_1+C_2}\cos\alpha_1 \omega^3_3+ e^{C_1+2C_2}\sin\alpha_1 \omega^4_3,
\end{align}
This time we see only the invariant forms of SO(3)$_{\text{D}}$ appearing, which is to be expected as we have already established that $\chi^4_{1,2}$ are singlets under the diagonal SO(3). This time we have that the entire of $(d-H\wedge)(e^{2A-\Phi}\Psi_{+})-2  e^{A-\Phi}\Psi_{-}$  gives rise to components of the flux (see  \eqref{eq:susycond7d}), which leaves us with a less constrained system of ODEs to solve. Plugging \eqref{eq:eq1bi} into  \eqref{eq:susycond7d} as before, we find the constraints
\begin{align}\label{eq:neq1zeroforms}
\Lambda_1(\alpha_1)&=\Lambda_2(\alpha_1)=0,\nn\\[2mm]
\Lambda_1(\alpha)&=e^{C_1+C_2}\cos\alpha(1+ 2 \lambda)+ (e^{2C_1}\lambda(1+ \lambda)-e^{2C_2})\sin\alpha,\nn\\[2mm]
\Lambda_2(\alpha)&= e^{3C_1}\sin\alpha(c_2+ 3 c_1 \lambda^2+2 c_1 \lambda^3)+ c_1e^{C_2} (3 e^{2C_1}\lambda(1+\lambda) + e^{2C_2})\cos\alpha,
\end{align}
as well as four ODEs
\begin{align}\label{eq:susyneq1}
&\partial_{\rho}(e^{2A+3C_1-\Phi}\sin\alpha_1)-c_1 e^{2A+k-\Phi}\sin\alpha_2=0,\nn\\[2mm]
&\partial_{\rho}\big(e^{2A-\Phi}(e^{2C_1+ C_2}\cos\alpha_1+ e^{3C_1} \lambda \sin\alpha_1)\big)- e^{2A+k-\Phi}(2e^{C_1+ C_2}\cos\alpha_2+ \sin\alpha_2 (p+ c_1 \lambda))=0,\nn\\[2mm]
&\partial_{\rho}\big(e^{2A-\Phi}(-e^{3C_2}\cos\alpha_1-3 e^{C_1+ 2C_2}\lambda \sin\alpha_1+ 3 e^{2C_1+C_2}\lambda^2+ e^{3C_1}\lambda^3\sin\alpha_1)\big)- c_2 e^{2A+k-\Phi}\sin\alpha_2=0,\nn\\[2mm]
& e^{2(C_1+C_2)}\left(\frac{3}{2} e^{A}\partial_{\rho}(p+ c_1 \lambda)+ 2 e^{C_1+C_2}(3 e^k- e^{A} \partial_{\rho}\alpha_2)\right)+e^{A+k}\cos\alpha_2\Lambda_2\left(\alpha_1-\frac{\pi}{2}\right)\nn\\
&+3e^{A+C_1+k}\big(p \cos\alpha_2-2 e^{C_1+C_2}\sin\alpha_2\big)\Lambda_1\left(\alpha_1-\frac{\pi}{2}\right)=0.
\end{align}
The Romans mass on the other hand are defined as
\begin{align}
&e^{3A+3C_1+3C_2+k} F_0+\partial_{\rho}(e^{3(A+C_1+C_2)-\Phi}\sin\alpha_2)-\frac{3}{2}e^{3A+2(C_1+C_2)-\Phi}\cos\alpha_2\partial_{\rho}(p+ c_1 \lambda)\nn\\[2mm]
&+ e^{2A-\Phi+k}\bigg[e^{A+C_1}\cos\alpha_1\big(-3 e^{2C_2}p+ e^{2C_1}(c_2+ 3c_1 \lambda^2+ 2c_1\lambda^3+ 3p \lambda(1+ \lambda))\big)\\[2mm]
&- e^{C_2}\big(2 e^{3C_1+ 2C_2}\cos\alpha_2+ e^{A} \sin\alpha_1(c_1 e^{2C_2} + 3 e^{2C_1}(c_1 \lambda(1+ \lambda)+ p(1+ 2\lambda)))\big)\bigg]=0\nn
\end{align}
with the remaining RR fluxes defined as in \eqref{eq:RRansatz} for specific functions
\begin{align}
e^{3A+C_1+C_2+k}u_1&= \partial_{\rho}(e^{3A+2(C_1+C_2)+k-\Phi}\cos\alpha_2)+e^{3A+C_1+C_2-\Phi}\sin\alpha_2\partial_{\rho}(p+ c_1 \lambda)\nn\\[2mm]
&+ e^{2A+C_1+k-\Phi}\big(e^{A}\cos\alpha_1(e^{2C_1}\lambda(1+ \lambda)- e^{2C_2})+ e^{C_1+C_2}(e^{C_2} \sin\alpha_2- e^{A}\sin\alpha_1(1+ 2\lambda))\big),\nn\\[2mm]
e^{A+3C_2}u_2&=e^{A+3C_1+k-\Phi}\big(c_2+ 3 c_1\lambda^2+ 2 c_1\lambda^3+ 3 p \lambda(1+ \lambda)\big)\cos\alpha_2\nn\\[2mm]
&-2 e^{C_2+3C_1+k-\Phi}\big(e^{2C_2}\cos\alpha_1+3 e^{A+C_1}\lambda(1+ \lambda)\sin\alpha_2\big),\nn\\[2mm]
e^{A+3C_1}u_3&=-e^{3C_2+k-\Phi}(e^{A}c_1 \cos\alpha_2+ 2 e^{3C_1}\sin\alpha_1),\nn\\[2mm]
e^{A+C_2}u_4&=2 e^{2C_1+C_2+k-\Phi}\big(e^{C_2}\sin\alpha_2- e^{A}(1+ 2\lambda)\sin\alpha_2\big)+ e^{A+C_1+k-\Phi}\big(c_1\lambda(1+\lambda)+ p (1+2 \lambda)\big),\nn\\[2mm]
e^{A+C_1}u_5&= 2 e^{2(C_1+C_2)+k-\Phi}\cos\alpha_1+ e^{A+C_2+k-\Phi}(p\cos\alpha_2-2 e^{C_1+C_2}\sin\alpha_2),\nn\\[2mm]
e^{3A+k}u_6&=-e^{C_1+C_2}\bigg(\partial_{\rho}(e^{3A+C_1+C_2-\Phi}\sin\alpha_2)+\frac{1}{2}e^{2A-\Phi}\cos\alpha_2\big(4 e^{C_1+C_2+k}+ e^{A}\partial_{\rho}(p+ c_1 \lambda)\big)\bigg),\nn\\[2mm]
e^{3A+k}u_7&= e^{3(C_1+C_2)}\big(\partial_{\rho}(e^{3A-\Phi}\cos\alpha_2)+ 2 e^{2A+k-\Phi}\sin\alpha_2\big).
\end{align}
Plugging these functions into the conditions that imply the RR Bianchi identities we find that, unlike the $\mathcal{N}=(3,0)$ case, in general none are implied automatically by \eqref{eq:susyneq1}. This should be expected for $\mathcal{N}=(1,0)$ as completely general Bianchi identities are generically rather long and unwieldy unless they happen to be implied by extended supersymmetry. Nonetheless, in section \ref{sec:exneq1} we are able to solve all the necessary conditions for supersymmetry and the Bianchi identities for some sub cases within this class.\\
~\\
 In the next section we shall present some new solutions preserving $\mathcal{N}=(3,0)$ supersymmetry. 
\section{$\mathcal{N}=(3,0)$ solutions in massive IIA}\label{sec:exneq3}
In this section we present some analytic and series solutions we have found to the system of ODEs that implies $\mathcal{N}=(3,0)$ supersymmetry in section \ref{eq:condtionneq3}. For the series solutions, we show that it is possible to interpolate numerically between these behaviours leading to a compact internal space. We will consider 3 cases in the next 3 subsections, $c=F_0=0$, $c=0$ and $F_0=0$.

\subsection{Closed form solution: $c=F_0=0$}
In the case $c=F_0=0$ it is possible to use \eqref{eq:ODE-system1} to define the dilaton in terms of the other functions:
\begin{equation}
\label{eq:analytic_dildef}
e^{\Phi} = g e^A \tan \alpha_1 \sin \alpha_2 
\end{equation}
where $g$ is a constant. Substituting this expression for $\Phi$ back into the ODE system and by fixing the reparametrisation invariance so that $k = -A-\log(4 \cos \alpha_1 \cos \alpha_2)$ we are left with the following equations:
\begin{align}
&\partial_\rho(e^{2A} \sin \alpha_2 \sec \alpha_1) = 0, \label{eq:analyticODE1} \\[2mm]
&\partial_\rho (e^{2A} \tan \alpha_2) + \frac{(\cot^2 \alpha_1 -1)(1+\sec^2 \alpha_2)}{4 \cos \alpha_1 \cos \alpha_2}=0, \label{eq:analyticODE2}\\[2mm]
&\partial_\rho ( \cos (2 \alpha_1))-e^{-2A}\cos \alpha_1 \csc \alpha_2=0\label{eq:analyticODE3} .
\end{align}
Now it is immediate to notice that by deriving \eqref{eq:analyticODE3} and using \eqref{eq:analyticODE1} we can determine $\alpha_1$ as a function of $\rho$
\begin{equation}
\alpha_1 = \frac{1}{2} \arccos (a\rho+b).
\end{equation}
Using parametrization invariance respect to dilatation and translation we also map $a \rho + b \to \rho$. One can then plug this expression back to \eqref{eq:analyticODE3} and find $A$ as a function of $\alpha_2$ and finally use \eqref{eq:analyticODE2} to get an explicit solution for $\alpha_2$, which reads:
\begin{equation}
\alpha_2 = \arccos \left( \sqrt{\frac{2(-\rho+\kappa+\log(1+\rho))}{-1+\rho}} \right) ,
\end{equation}
where $\kappa$ is the integration constant.
After all these manipulations, the explicit expressions for the warping functions are:
\begin{align}
&e^{2A} = \frac{1}{a} \sqrt{\frac{1-\rho^2}{2(1+2\kappa-3\rho+2\log(1+\rho))}}, \qquad e^{2k} =  \frac{1}{a}  \sqrt{\frac{(1-\rho)(1+2\kappa-3\rho+2\log(1+\rho))}{128(1+\rho)^3 (\kappa-\rho+\log(1+\rho))^2}} , \nn \\[2mm]
&e^{2C_1} =  \frac{1}{a}  \sqrt{\frac{2(1+2\kappa-3\rho+2\log(1+\rho))}{1-\rho^2}}, \qquad e^{2C_1} = \frac{1}{a}  \sqrt{2(1-\rho^2)(1+2\kappa-3\rho+2\log(1+\rho))}.
\end{align}

\begin{figure}
	\centering
	\includegraphics[scale=0.6]{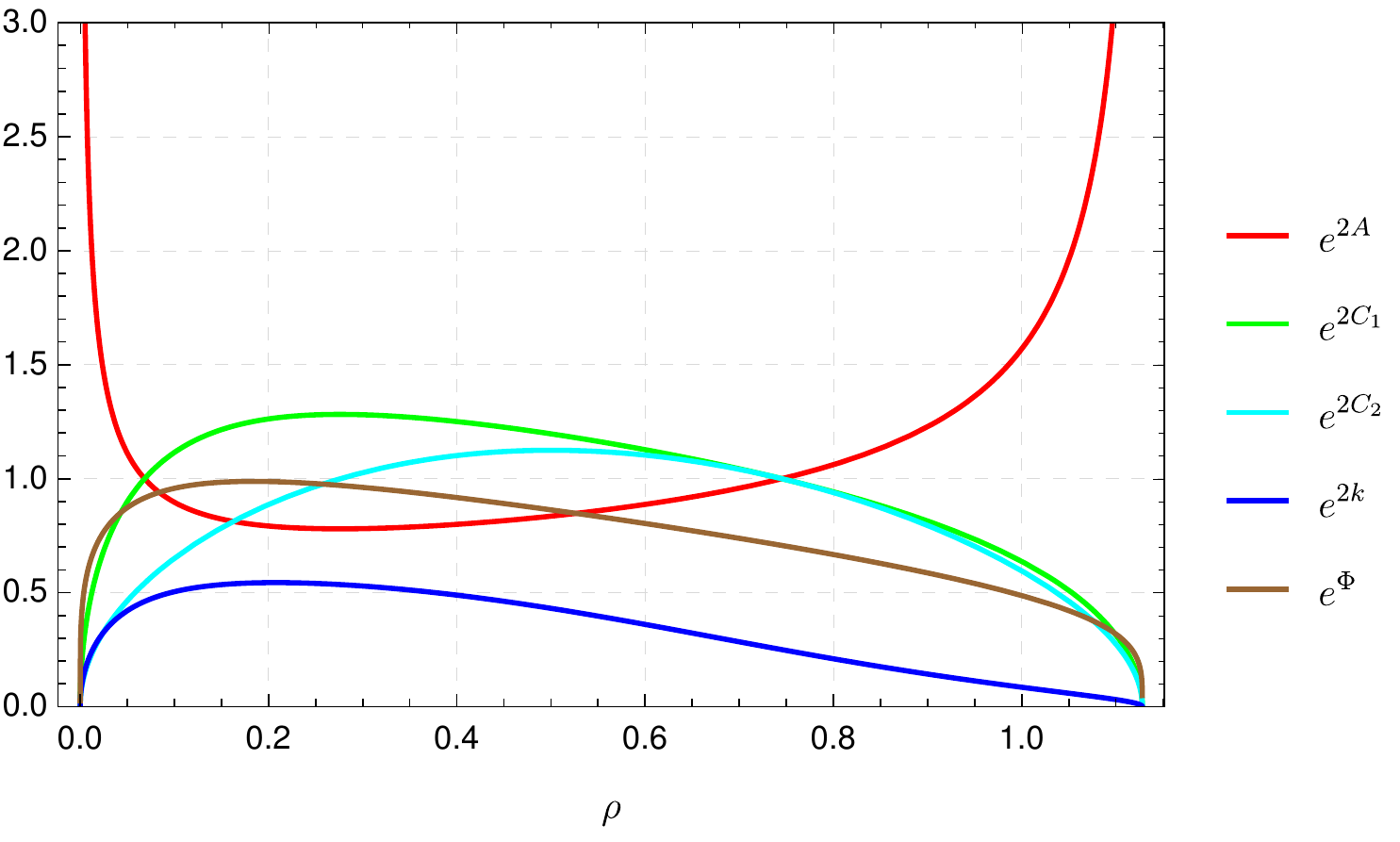}
	\caption{Behaviour of warping factors and dilaton for $\kappa=-1/4$ and $a=1$ as functions of $\rho$ for the O2-O2 system. The solution has been shifted from $\rho$ to $\rho - \rho_1$, so that the first O2-plane is at $\rho=0$.}
	\label{fig:O2-O2}
\end{figure}

Even though this solution admits various possible realizations depending on the value of $\kappa$ - none of them give rise to regular compact solution. There does however exist values of $\kappa$ for which the solution is bounded by  physical singularities. In particular, when $k$ lives in the interval $(\log(3/2)-1,0)$, $\rho$ is bounded between the two solutions of the equation $1+2\kappa-3\rho+2\log(1+\rho)=0$ we denote $\rho_{1,2}$. Since $\rho_{1,2}$ are first order poles it is easy to see that they each give rise to the behaviour of O2-plane that wrap AdS$_3$ at these loci, i.e. $e^{2 C_{1,2}},e^{2k} \sim |\rho-\rho_{1,2}|^\frac{1}{2}$ while $e^{2A} \sim |\rho-\rho_{1,2}|^{-\frac{1}{2}}$. An example of such O2-O2 system with $\kappa=-1/4$ is given in figure \ref{fig:O2-O2}. This interpretation is also confirmed by the dilaton, which can be checked using \eqref{eq:analytic_dildef} goes as $e^\Phi \sim |\rho-\rho_{1,2}|^\frac{1}{4}$.

\subsection{Some solutions with $c=0$}

Let's now consider the slightly more difficult case where we have $c=0$ but non-vanishing Romans mass. Again, thanks to \eqref{eq:ODE-system2} the dilaton is determined as
\begin{equation}
e^\Phi = g \frac{e^{5A} \sin^3 \alpha_2 \sin \alpha_1}{\cos^3 \alpha_1}
\end{equation}
and we can reduce the problem of finding a solution to a system of 3 ODEs. This time however we will limit ourselves to a perturbative analysis near to a given physical singularity. 

Let's start by considering an O2-like behaviour. Thanks to translation invariance, we can assume without loss of generality that the O2 sits at $\rho=0$. As one can check by the definition of the warp functions \eqref{eq:neq3defs}, this is given by the following expansion of the functions
\begin{equation}
\label{eq:O2_exp}
\sin^2 \alpha_1 = k_1 + k_2 \rho + k_3 \rho^2 + O(\rho^3), \quad \sin^2 \alpha_2 = h_1 \rho + h_2 \rho^2 + O(\rho^3), \quad e^{4A} = \frac{a_1+a_2 \rho}{\rho} + O(\rho) ,
\end{equation}  
\begin{figure}
	\centering
	\includegraphics[scale=0.6]{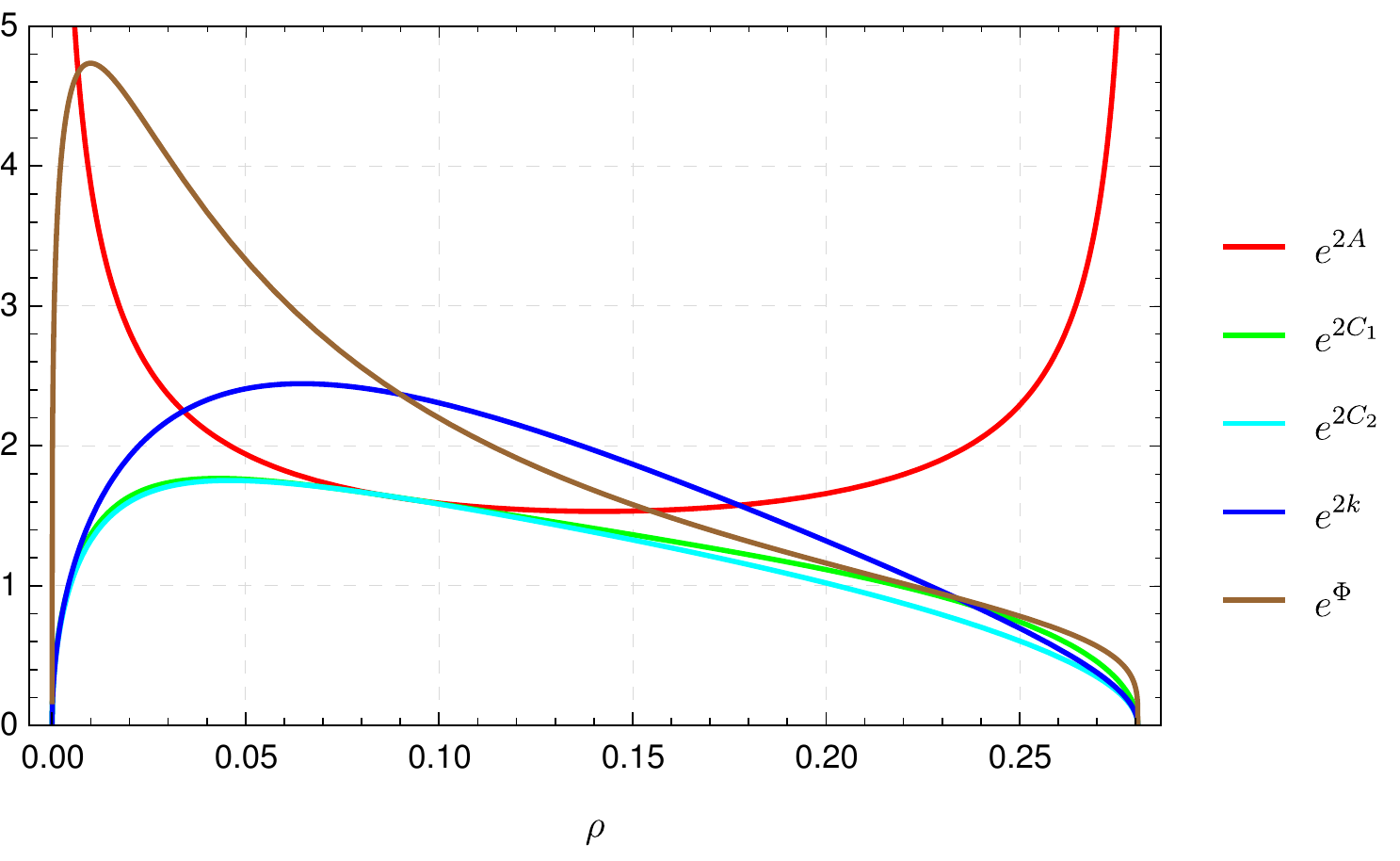}
	\caption{Behaviour of warping factors and dilaton for $F_0=1$ as functions of $\rho$. The input data for the numerics are $k_1=0.6$ and $k_2^2=$1.536. At both the endpoints the solution behaves like an O2 plane wrapped on AdS$_3$.}
	\label{fig:O2-O2_c}
\end{figure}
where $(k_1,h_1,a_1)$ must be non vanishing - we also choose to fix $k=-A+\log(2\tan\alpha_1)$. The supersymmetry conditions \eqref{eq:ODE-system1}-\eqref{eq:ODE-system4}  impose that the coefficients of the series expansion are all determined in terms of $k_{1,2}$ as following:
\begin{align}
&h_1 =\frac{(3 k_1-2) k_2^3-32 F_0 k_1^3}{(k_1-1) k_1 k_2^2} , \quad k_3 = \frac{(1-2 k_1) k_2^2}{4 (k_1-1) k_1}, \nn \\[2mm]
&h_2 = \frac{-4096 F_0^2 k_1^6+128 F_0 (2 k_1-1) k_1^3 k_2^3+\left(-18 k_1^2+25 k_1-10\right) k_2^6}{4 (k_1-1)^2 k_1^2 k_2^4} , \\[2mm]
&a_1 = \frac{16 (k_1-1)^4}{32 F_0 k_1^3+(2-3 k_1) k_2^3} , \quad a_2 = \frac{4 (k_1-1) k_1 k_2 \left(128 F_0 k_1^3+\left(-30 k_1^2+33 k_1-10\right) k_2^3\right)}{\left(32 F_0 k_1^3+(2-3 k_1) k_2^3\right)^2} . \nn
\end{align}
This series expansion can be used to run a numerical analysis starting in the proximity of $\rho=0$. The result is given in figure \ref{fig:O2-O2_c}. We find  that for this solution $\rho$ is actually bounded and on the other side warping functions and dilaton are compatible with another O2. Thus we find an numerical solution bounded between two O2 planes, the functions appearing in the solution are plotted in figure \ref{fig:O2-O2_c}.

\begin{figure}
	\centering
	\includegraphics[scale=0.6]{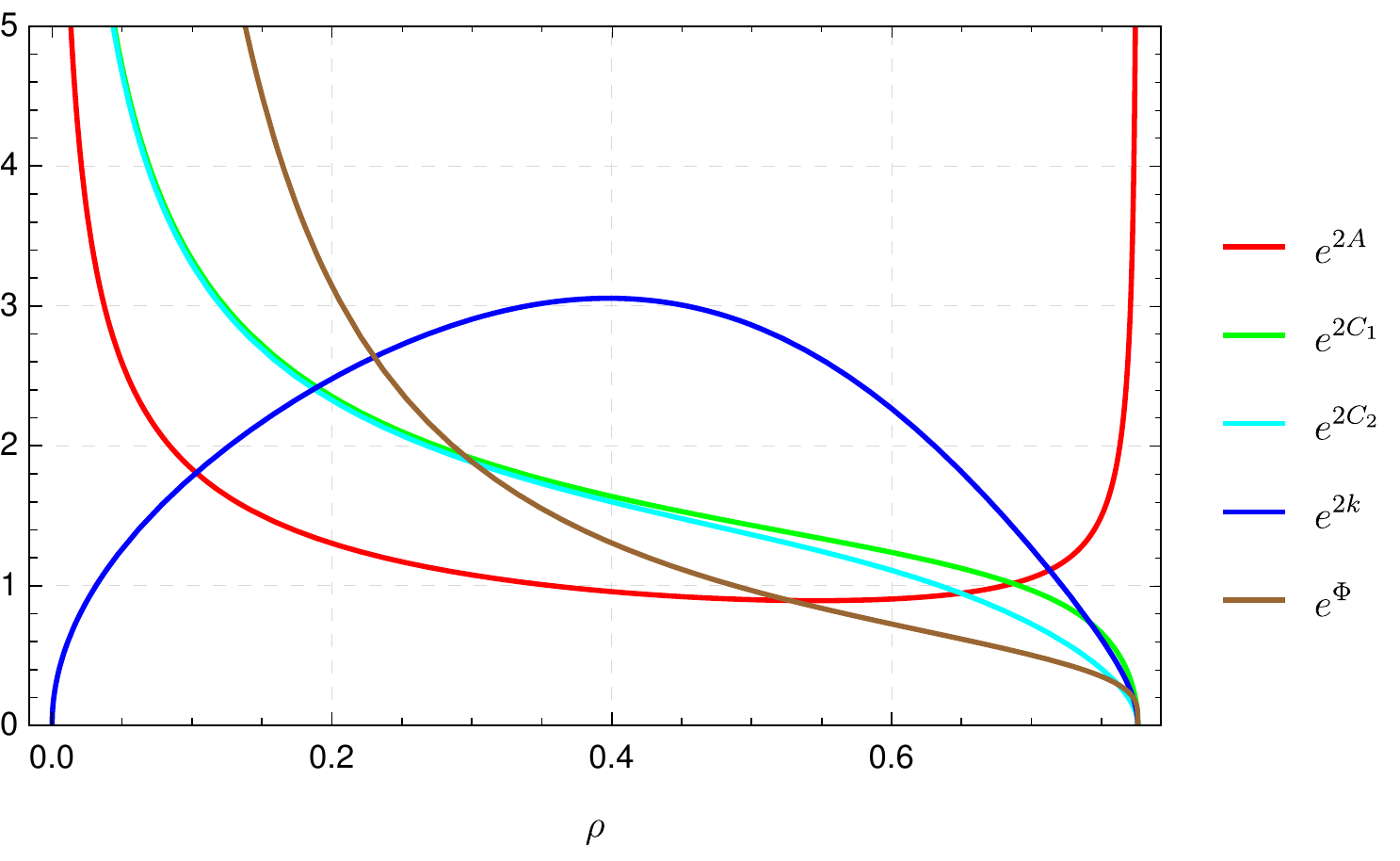}
	\caption{Behaviour of warping factors and dilaton for $F_0=1$ as functions of $\rho$. The input data for the numerics is $k_1=0.45$. At $\rho=0$ the functions behaves like an O8, while at the endpoint it behaves like an O2.}
	\label{fig:O2-O8_c}
\end{figure}
Now we would like to recover  D8/O8-like behaviour, which near the singularity is given by $e^{2k} \sim \rho^\frac{1}{2}$, $e^{2A},e^{2 C_{1,2}} \sim \rho^{-\frac{1}{2}}$ while the dilaton goes like $e^\Phi \sim \rho^{-\frac{5}{4}}$. In this case we already know that such a solution should exist since we can fix the fibration $\lambda=0$ and recover the local $\mathcal{N}=(4,0)$ solution with D8/O8s sources in \cite{Macpherson:2018mif}. Let us see if we can find a deformation of that particular case preserving just $\mathcal{N}=(3,0)$. Again we proceed with a series expansion, however in this case the first non-zero coefficients appear at higher order with respect to the previous case:
\begin{equation}
\sin^2 \alpha_1 = k_1 + k_2 \rho^3 + O(\rho^4), \quad \sin^2 \alpha_2 = 1 + h_2 \rho^3 + O(\rho^4), \quad e^{4A} = \frac{a_1+a_2 \rho^3}{\rho} + O(\rho^3).
\end{equation}  
Notice that by setting $k_1 = \frac{1}{2}, a_1 \sim F_0^{-1}$ and everything else to zero we would fall to the case studied in  \cite{Macpherson:2018mif}. However it can be showed that the ODE system admits also a solution with
\begin{align}
&k_2 = \frac{8 F_0 k_1 (2 k_1-1)}{3 (k_1-1)^2} , \quad \, \, h_2 = \frac{2 F_0 ((1 - 2 k_1)^2)}{(-1 + k_1)^3} , \\
&a_1 = \frac{(k_1-1)^2}{2 F_0 k_1} , \qquad \qquad  a_2 = \frac{20 k_1^2-96 k_1+43}{24 k_1 \left(1-k_1\right)} .
\end{align} 
Remarkably, the numeric solution associated to this series expansion reproduces a compact behaviour, as showed in figure \ref{fig:O2-O8_c} where we fixed have $F_0=1$. We have checked that the second boundary behaviour is that of an O2-plane.

\subsection{Some solutions with $F_0=0$}
A similar analysis to the one performed in the previous section can be carried out also in the massless case where we now assume that $c \neq 0$. Again \eqref{eq:ODE-system1} fixes the dilaton to be as in equation \eqref{eq:analytic_dildef} and we are left with a system of three ODEs to solve.
Since Romans mass is turned of, in this section we will be mostly interested in O2 solutions, and therefore we will look for an expansion like the one in \eqref{eq:O2_exp} 
\begin{equation}
\cos^2 \alpha_1 = k_1 + k_2 \rho + k_3 \rho^2 + O(\rho^3), \quad \sin^2 \alpha_2 = h_1 \rho + h_2 \rho^2 + O(\rho^3), \quad e^{-4A} = a_1 \rho +a_2 \rho^2 + O(\rho^3) ,
\end{equation}  
though here, for convenience, we have rearranged the functions in a slightly different (but equivalent) way. We also fix $k= 3A + \log(\sin^2 \alpha_2 \tan \alpha_1)$.

\begin{figure}
	\centering
	\subfigure[Numerical solution associated to \eqref{eq:expansion1}. $\qquad$ Input data: $c=1,b=0.13$]{\includegraphics[scale=0.6]{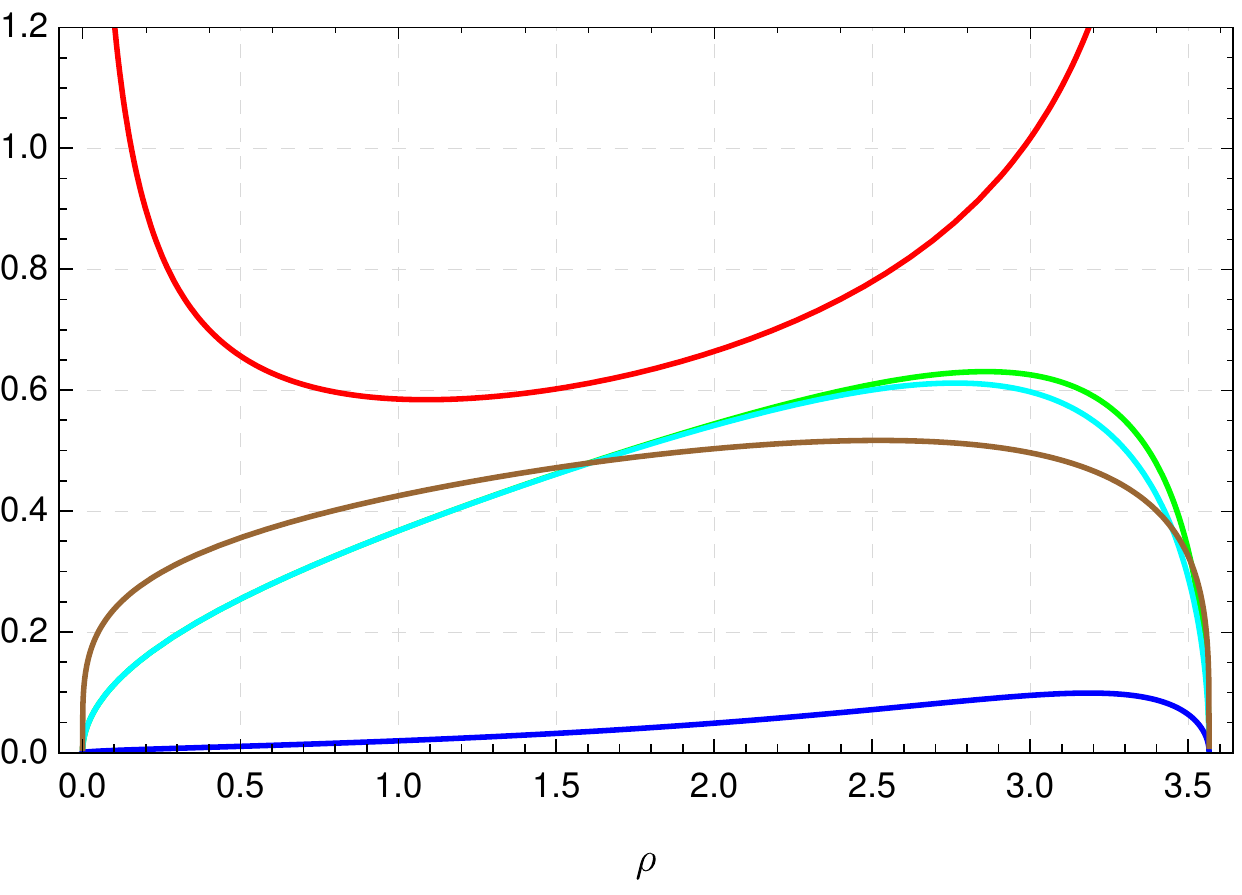}\label{fig:O2-O8_m}}
	\subfigure[Numerical solution associated to \eqref{eq:expansion2}. $\qquad$ $\qquad$ $\qquad$ Input data: $c=2,b=1,k_1=0.9$]{\includegraphics[scale=0.6]{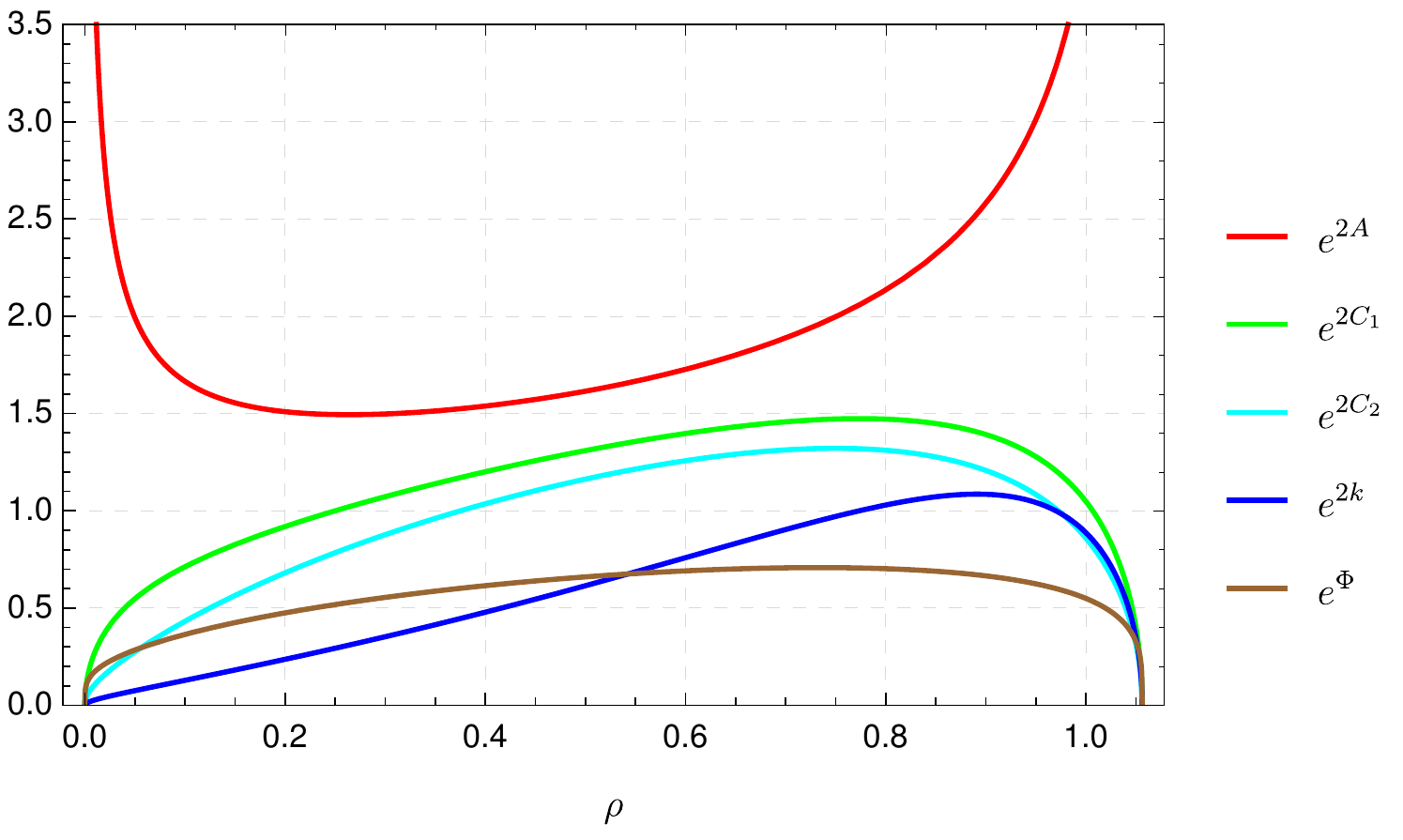}\label{fig:O2-O8_m2}}
	\caption{Behaviour of warping factors and dilaton as functions of $\rho$. In both cases the solution is an O2-O2 system.}
\end{figure}

By inserting this expansion inside the system ODEs  we find that there are two possible expansion compatible with an O2 singularity at $\rho =0$. For the first one the ODE system imposes:
\begin{align}
&k_1 = \frac{1}{2}, \quad k_2 =\frac{\sqrt{b}}{\sqrt{2}}-\frac{c}{4} , \quad k_3 =\frac{1}{16} c \left(c-2 \sqrt{2} \sqrt{b}\right) ,\nn \\
&h_1 = 3 \sqrt{2} \sqrt{b}-c , \quad h_2 = -\frac{7 c^2}{4}+\frac{19 c \sqrt{b}}{\sqrt{2}}-26 b , \label{eq:expansion1}\\
&a_1 = -\frac{2 \left(c-3 \sqrt{2} \sqrt{b}\right)}{b} , \quad a_2 = -\frac{7 c^2}{2 b}+\frac{21 \sqrt{2} c}{\sqrt{b}}-64 \nn\\
\end{align}
where we used the auxiliary parameter $b> 12 c$ for convenience. In order to simplify the expression we also assumed $b> 8 c$, even if it was not necessary in principle.
The second solution is given by
\begin{align}
&k_2 = k_1 \left(2 \sqrt{b-b k_1}-c k_1\right) , \quad k_3=k_1 \left(c k_1 \left(c k_1-3 \sqrt{b-b k_1}\right)+b\right) , \nn \\[2mm]
&a_1 = \frac{c}{b-b k_1}+\frac{2-6 k_1}{k_1 \sqrt{b-b k_1}} , \quad h_1 =\frac{2 (3 k_1-1) \sqrt{b-b k_1}-c k_1}{k_1-1} , \nn \\[2mm]
&a_2 = \frac{b \left(42 k_1^3-79 k_1^2+46 k_1-9\right)-c k_1 \left((4-11 k_1) \sqrt{b-b k_1}+c k_1\right)}{b (k_1-1)^2 k_1} , \nn \\[2mm]
&h_2 = \frac{b (k_1-1) (3 k_1 (10 k_1-7)+5)-c k_1 \left((2-9 k_1) \sqrt{b-b k_1}+c k_1\right)}{(k_1-1)^2}, \label{eq:expansion2}
\end{align}
where again we introduced the parameter $b$.

Running a numerical analysis on both these expansions, we found that the interpolating functions have $\rho$ bounded in an interval and in both cases on the other endpoint they display an O2-like behaviour. The situation for both the cases considered in this section is displayed in figure \ref{fig:O2-O8_m} for the first expansion and \ref{fig:O2-O8_m2} for the second one.

\section{$\mathcal{N}=(1,0)$ solutions in massive IIA}\label{sec:exneq1}

In this section we present some simple closed form solutions that lie within the class of section \ref{sec:neq1conds}. These are by no means exhaustive, in fact they are essentially the most simple ways to solve the necessary conditions to have a solution and still end up with just $\mathcal{N}=(1,0)$ supersymmetry. 

In section \ref{sec:neq1ex1} we first explore the possibility of solutions with $\lambda=0$, unlike the $\mathcal{N}=(3,0)$ case, one such solution exists that is not enhancement to $\mathcal{N}=(4,0)$. Then in sections \ref{sec:neq1ex2} and \ref{sec:neq1ex3} we present unique solutions that follow from solving the necessary condition (see \eqref{eq:neq1zeroforms})
\beq\label{eq:cond}
e^{C_1+C_2}\cos\alpha_1(1+ 2 \lambda)+ \sin\alpha_1(e^{2C_1}\lambda(1+ \lambda)-e^{2C_2})=0
\eeq
as respectively
\beq
\cos\alpha_1= e^{2C_1}\lambda(1+ \lambda)-e^{2C_2}=0,~~~~~\text{and},~~~~~\lambda+1= e^{C_1}\cos\alpha_1+ e^{C_2}\sin\alpha_1=0.
\eeq
Another simple way to solve \eqref{eq:cond} is to fix $1+2\lambda= \sin\alpha_1=0$. However we do not believe that any solution of this kind exists\footnote{We have confirmed that no such solutions exist when we additionally assume $\sin\alpha_2=0$ - turning on this function does not appear to improve matters, though we have not rigorously ruled out the possibility of such solutions.} and that further solutions beyond what we present here must solve \eqref{eq:cond} in a generic manner, and likely come with non-constant $\lambda$.\\
~~\\
In the following subsections we present several closed form solutions that are unique for these tunings of $\lambda$ and $\cos\alpha_1=0$. Confirming that these solve the supersymmetry conditions and Bianchi identities of section \ref{sec:neq1conds} and \eqref{eq:bis} is not hard - proving that they are unique for these tunings is rather laborious. We will spare the details and just present the result.
\subsection{Unique solution with $\lambda=0$}\label{sec:neq1ex1}
Fixing $\lambda=0$ means that we non longer have a S$^3\times$S$^3$ fibration, and it is just via the fluxes that supersymmetry is broken to $\mathcal{N}=(1,0)$. Plugging $\lambda=0$ into the conditions  \eqref{eq:susyneq1} and \eqref{eq:neq1zeroforms}, fixes many of the function - then plugging these into the Bianchi identities fixes yet more. After considerable manipulation one arrives at an analytic expression for the functions of our ansatz which reduces to a unique solution of closed form. This is defined by the following expressions
\begin{align}
e^{A}&= L\frac{\cos^{\frac{1}{6}} \rho  }{\sqrt{\sin \rho}},~~~ e^{C_1}=\frac{L \cos^{\frac{1}{6}} \rho\sqrt{\sin \rho}}{\cos\alpha_1},~~~e^{C_2}=\frac{L \cos^{\frac{1}{6}} \rho\sqrt{\sin \rho}}{\sin\alpha_1},\nn\\[2mm]
 e^{-\Phi}&= \frac{F_0 L}{2\cos^{\frac{5}{6}} \rho \sqrt{\sin \rho}},~~~ e^k= L\frac{\sin^{\frac{3}{2}} \rho}{\cos^{\frac{11}{6}} \rho},~~~ \lambda = 0
\end{align}
for $L$ a constant. One additionally needs to fix
\beq
c_1=c_2=0,~~~~ \alpha_2=\rho,~~~\partial_{\rho} \alpha_1=0,~~~~ p= - \frac{2L^2 \cos^{\frac{4}{3}} \rho}{\sin\alpha_1\cos\alpha_1},
\eeq
from which all the bosonic fields follow. One can check that close to $\rho=0$ the metric and dilaton behave as an O2 plane wrapped on AdS$_3$, however at $\rho=\frac{\pi}{2}$ the behaviour is the one far from a D8-brane. This is located at infinite proper distance from the origin - so in this case the internal space is non compact. It may be possible to resolve this issue, by gluing local solutions together with D8-branes on the interior of the interval, but we will not attempt to do this here. 

\subsection{Unique solution with $\lambda=-1$}\label{sec:neq1ex2}
From the point of view of the metric,  fixing $\lambda=-1$ is equivalent to setting $\lambda=0$, as the former is a gauge transformation (within the SU(2) invariant forms of the 3-spheres) of the later. This however is not true for all of the SO(3)$_{\text{D}}$ invariant forms, so $\lambda=-1$ can potentially give rise to a distinct solution. Following the procedure sketch in the previous section on finds
\begin{align}
e^{A}&= -L\frac{\cos^{\frac{1}{6}} \rho  }{\sqrt{\sin \rho}},~~~ e^{C_1}=-\frac{L \cos^{\frac{1}{6}} \rho\sqrt{\sin \rho}}{\cos\alpha_1},~~~e^{C_2}=\frac{L \cos^{\frac{1}{6}} \rho\sqrt{\sin \rho}}{\sin\alpha_1},\nn\\[2mm]
 e^{-\Phi}&= \frac{F_0 L}{2\cos^{\frac{5}{6}} \rho \sqrt{\sin \rho}},~~~ e^k= -L\frac{\sin^{\frac{3}{2}} \rho}{\cos^{\frac{11}{6}} \rho},
\end{align}
and
\beq
c_1=c_2=0,~~~~ \alpha_2=\rho,~~~\partial_{\rho}\alpha_1=0,~~~~ p=  \frac{2L^2 \cos^{\frac{4}{3}} \rho}{\sin\alpha_1\cos\alpha_1},
\eeq
Notice that this solution is essentially equal to the one of the previous section up to some signs, so physically this solution is indistinguishable from that one. 

\subsection{Unique solution with $\cos\alpha_1=0$}\label{sec:neq1ex3}
The final solution we consider in this section is also the more interesting. Fixing $\cos\alpha_1=0$ and (without loss of generality) $\sin\alpha_1=1$ eventually fixes $\lambda=\lambda_0=$ constant -  however the fibration is strictly not topologically trivial. We find for the metric fields, AdS warp factor and dilaton
\begin{align}
e^{A}&= -\frac{L\Delta^{\frac{1}{6}}}{\cos^{\frac{1}{6}} \rho \sqrt{\sin \rho}},~~~e^{C_1}=\frac{L \Delta^{\frac{1}{6}}(1+ 2\lambda_0)\sqrt{\sin\rho}}{\sqrt{\lambda_0(1+\lambda_0)}\cos^{\frac{1}{6}}\rho} ,~~~e^{C_2}=-\frac{L \Delta^{\frac{1}{6}}(1+ 2\lambda_0)\sqrt{\sin\rho}}{\cos^{\frac{1}{6}}\rho},\nn\\[2mm]
e^{k}&= -\frac{2L (1+ 2\lambda_0)^2 \sin^{\frac{3}{2}} \rho}{3 \Delta^{\frac{5}{6}}\cos^{\frac{1}{6}}\rho},~~~ e^{-\Phi}= \frac{F_0 L (1+ 2 \lambda_0)^2 \cos^{\frac{5}{6}}\rho}{\Delta^{\frac{5}{6}}\sqrt{\sin \rho}},
\end{align}
where again $L$ is a constant and we have introduced 
\beq
\Delta= 1+ 8 \lambda_0(1+ \lambda_0)+(1+ 2\lambda_0)^2\cos2\rho.
\eeq
The remaining functions and constants are fixed to
\beq
c_1=c_2=0,~~~~,\alpha_2= \rho,~~~~ p= \frac{2 L^2 \Delta^{\frac{1}{3}}(1+ 2\lambda_0)^2\cos^{\frac{2}{3}}\rho}{\sqrt{\lambda_0(1+ \lambda_0)}}.
\eeq
As with the solutions in the previous two sections, close to $\rho=0$ the metric and dilaton reproduce the behavior of O2-branes wrapped AdS$_3$. This time however, the behavior close to $\rho=\frac{\pi}{2}$ is that of a D8/O8 system wrapped on AdS$_3\times$S$^3_1\times$S$^3_2$. As such the internal space is bounded and so this solution is a viable candidate for a holographic dual.\\
~\\
It would be interesting to study this solution in more detail, and indeed to find what else lurks within our $\mathcal{N}=(1,0)$ system - we leave this for future work.

\section*{Acknowledgments}
We thanks Alessandro Tomasiello for useful discussion. AL is supported by INFN and by MIUR-PRIN contract 2017CC72MK003.


\appendix

\section{Hodge dual of the SO(3)$_{\text{D}}$ invariant forms}\label{hodge}
In this appendix we present identities relating the various SO(3)$_{\text{D}}$ invariant forms under the seven-dimensional Hodge dual, namely
\begin{align}
\star_7 1&= e^{3(C_1+C_2)+k} d \rho\wedge\omega^1_3\wedge \omega^2_3,\nn\\[2mm]
\star_7 \omega_2&= e^{C_1+C_2+k}d \rho\wedge \omega_4,\nn\\[2mm]
\star_7 e^{3C_1}\omega^1_3&=-e^{3C_2+k}d \rho\wedge\omega^2_3,\nn\\[2mm]
\star_7 e^{3C_2}\omega^2_3&=e^{3C_1+k}d \rho\wedge\omega^1_3,\nn\\[2mm]
\star_7 e^{C_1}\omega^3_3&=e^{C_2+k}d \rho\wedge\omega^4_3,\nn\\[2mm]
\star_7 e^{C_2}\omega^4_3&=e^{C_1+k}d \rho\wedge\omega^3_3,\nn\\[2mm]
\star_7 e^{C_1+C_2} \omega_4&= e^{k}d \rho\wedge \omega_2,\nn\\[2mm]
\star_7e^{3(C_1+C_2)} \omega^1_3\wedge \omega^2_3&=e^{k}d \rho.
\end{align}


\begin{thebibliography}{99}



\bibitem{Maldacena:2000hw}
  J.~M.~Maldacena and H.~Ooguri,
  ``Strings in AdS(3) and SL(2,R) WZW model 1.: The Spectrum,''
  J.\ Math.\ Phys.\  {\bf 42} (2001) 2929
  doi:10.1063/1.1377273
  [hep-th/0001053].
	
\bibitem{Maldacena:2000kv}
  J.~M.~Maldacena, H.~Ooguri and J.~Son,
  ``Strings in AdS(3) and the SL(2,R) WZW model. Part 2. Euclidean black hole,''
  J.\ Math.\ Phys.\  {\bf 42} (2001) 2961
  doi:10.1063/1.1377039
  [hep-th/0005183].
	
\bibitem{Maldacena:2001km}
  J.~M.~Maldacena and H.~Ooguri,
  ``Strings in AdS(3) and the SL(2,R) WZW model. Part 3. Correlation functions,''
  Phys.\ Rev.\ D {\bf 65} (2002) 106006
  doi:10.1103/PhysRevD.65.106006
  [hep-th/0111180].
	
\bibitem{Gaberdiel:2018rqv}
  M.~R.~Gaberdiel and R.~Gopakumar,
  ``Tensionless string spectra on AdS$_{3}$,''
  JHEP {\bf 1805} (2018) 085
  doi:10.1007/JHEP05(2018)085
  [arXiv:1803.04423 [hep-th]].
	

	
\bibitem{Eberhardt:2019ywk}
  L.~Eberhardt, M.~R.~Gaberdiel and R.~Gopakumar,
  ``Deriving the $\text{AdS}_{3}/\text{CFT}_{2}$ Correspondence,''
  arXiv:1911.00378 [hep-th].
	
\bibitem{Dei:2019osr}
  A.~Dei, L.~Eberhardt and M.~R.~Gaberdiel,
  ``Three-point functions in AdS$_{3}$/CFT$_{2}$ holography,''
  JHEP {\bf 1912} (2019) 012
  doi:10.1007/JHEP12(2019)012
  [arXiv:1907.13144 [hep-th]].
	
\bibitem{Eberhardt:2019qcl}
  L.~Eberhardt and M.~R.~Gaberdiel,
  ``String theory on $\boldsymbol{\text{AdS}_{\mathbf{3}}}$ and the symmetric orbifold of Liouville theory,''
  Nucl.\ Phys.\ B {\bf 948} (2019) 114774
  doi:10.1016/j.nuclphysb.2019.114774
  [arXiv:1903.00421 [hep-th]].

	
\bibitem{Martelli:2003ki}
  D.~Martelli and J.~Sparks,
  ``G structures, fluxes and calibrations in M theory,''
  Phys.\ Rev.\ D {\bf 68} (2003) 085014
  doi:10.1103/PhysRevD.68.085014
  [hep-th/0306225].

\bibitem{Kim:2005ez}
  N.~Kim,
  ``AdS(3) solutions of IIB supergravity from D3-branes,''
  JHEP {\bf 0601} (2006) 094
  doi:10.1088/1126-6708/2006/01/094
  [hep-th/0511029].
	


\bibitem{Gauntlett:2006af}
  J.~P.~Gauntlett, O.~A.~P.~Mac Conamhna, T.~Mateos and D.~Waldram,
  ``Supersymmetric AdS(3) solutions of type IIB supergravity,''
  Phys.\ Rev.\ Lett.\  {\bf 97} (2006) 171601
  doi:10.1103/PhysRevLett.97.171601
  [hep-th/0606221].
	
\bibitem{Gauntlett:2006ns}
  J.~P.~Gauntlett, N.~Kim and D.~Waldram,
  ``Supersymmetric AdS(3), AdS(2) and Bubble Solutions,''
  JHEP {\bf 0704} (2007) 005
  doi:10.1088/1126-6708/2007/04/005
  [hep-th/0612253].
	
\bibitem{Figueras:2007cn}
  P.~Figueras, O.~A.~P.~Mac Conamhna and E.~O Colgain,
  ``Global geometry of the supersymmetric AdS(3)/CFT(2) correspondence in M-theory,''
  Phys.\ Rev.\ D {\bf 76} (2007) 046007
  doi:10.1103/PhysRevD.76.046007
  [hep-th/0703275 [HEP-TH]].
	
\bibitem{Donos:2008hd}
  A.~Donos, J.~P.~Gauntlett and J.~Sparks,
  ``AdS(3) x (S**3 x S**3 x S**1) Solutions of Type IIB String Theory,''
  Class.\ Quant.\ Grav.\  {\bf 26} (2009) 065009
  doi:10.1088/0264-9381/26/6/065009
  [arXiv:0810.1379 [hep-th]].
	

	
\bibitem{DHoker:2008lup}
  E.~D'Hoker, J.~Estes, M.~Gutperle and D.~Krym,
  ``Exact Half-BPS Flux Solutions in M-theory. I: Local Solutions,''
  JHEP {\bf 0808} (2008) 028
  doi:10.1088/1126-6708/2008/08/028
  [arXiv:0806.0605 [hep-th]].
	
		
\bibitem{Colgain:2010wb}
  E.~O Colgain, J.~B.~Wu and H.~Yavartanoo,
  ``Supersymmetric AdS3 X S2 M-theory geometries with fluxes,''
  JHEP {\bf 1008} (2010) 114
  doi:10.1007/JHEP08(2010)114
  [arXiv:1005.4527 [hep-th]].
	
\bibitem{Estes:2012vm}
  J.~Estes, R.~Feldman and D.~Krym,
 ``Exact half-BPS flux solutions in $M$ theory with D(2,1;$c^\prime$;0)$^2$ symmetry: Local solutions,''
  Phys.\ Rev.\ D {\bf 87} (2013) no.4,  046008
  doi:10.1103/PhysRevD.87.046008
  [arXiv:1209.1845 [hep-th]].
	
\bibitem{Bachas:2013vza}
  C.~Bachas, E.~D'Hoker, J.~Estes and D.~Krym,
 ``M-theory Solutions Invariant under $D(2,1;\gamma) \oplus D(2,1;\gamma)$,''
  Fortsch.\ Phys.\  {\bf 62} (2014) 207
  doi:10.1002/prop.201300039
  [arXiv:1312.5477 [hep-th]].
	
	


	
	
\bibitem{Couzens:2017way}
  C.~Couzens, C.~Lawrie, D.~Martelli, S.~Schafer-Nameki and J.~M.~Wong,
  ``F-theory and AdS$_{3}$/CFT$_{2}$,''
  JHEP {\bf 1708} (2017) 043
  doi:10.1007/JHEP08(2017)043
  [arXiv:1705.04679 [hep-th]].
	
\bibitem{Couzens:2019wls}
  C.~Couzens, H.~h.~Lam, K.~Mayer and S.~Vandoren,
  ``Black Holes and (0,4) SCFTs from Type IIB on K3,''
  JHEP {\bf 1908} (2019) 043
  doi:10.1007/JHEP08(2019)043
  [arXiv:1904.05361 [hep-th]].
	
	
\bibitem{Couzens:2017nnr}
  C.~Couzens, D.~Martelli and S.~Schafer-Nameki,
  ``F-theory and AdS$_{3}$/CFT$_{2}$ (2, 0),''
  JHEP {\bf 1806} (2018) 008
  doi:10.1007/JHEP06(2018)008
  [arXiv:1712.07631 [hep-th]].
	
\bibitem{Couzens:2018wnk}
  C.~Couzens, J.~P.~Gauntlett, D.~Martelli and J.~Sparks,
  ``A geometric dual of $c$-extremization,''
  JHEP {\bf 1901} (2019) 212
  doi:10.1007/JHEP01(2019)212
  [arXiv:1810.11026 [hep-th]].
	
\bibitem{Couzens:2019iog}
  C.~Couzens,
  ``$\mathcal{N}=(0,2)$ AdS$_3$ Solutions of Type IIB and F-theory with Generic Fluxes,''
  arXiv:1911.04439 [hep-th].
	



	
\bibitem{Couzens:2019mkh}
  C.~Couzens, H.~h.~Lam and K.~Mayer,
  ``Twisted $\mathcal{N}=1$ SCFTs and their AdS$_3$ duals,''
  arXiv:1912.07605 [hep-th].
	
\bibitem{Passias:2019rga}
  A.~Passias and D.~Prins,
  ``On AdS$_3$ solutions of Type IIB,''
  arXiv:1910.06326 [hep-th].
	
\bibitem{Eberhardt:2017uup}
  L.~Eberhardt,
  ``Supersymmetric AdS$_{3}$ supergravity backgrounds and holography,''
  JHEP {\bf 1802} (2018) 087
  doi:10.1007/JHEP02(2018)087
  [arXiv:1710.09826 [hep-th]].
	
\bibitem{Kelekci:2016uqv}
 O.~Kelekci, Y.~Lozano, J.~Montero, E.~O.~Colgáin and M.~Park,
  ``Large superconformal near-horizons from M-theory,''
  Phys.\ Rev.\ D {\bf 93} (2016) no.8,  086010
  doi:10.1103/PhysRevD.93.086010
  [arXiv:1602.02802 [hep-th]].
	
	
\bibitem{Lozano:2019emq}
  Y.~Lozano, N.~T.~Macpherson, C.~Nunez and A.~Ramirez,
  ``AdS$_3$ solutions in Massive IIA with small $\mathcal{N}=(4,0)$ supersymmetry,''
  arXiv:1908.09851 [hep-th].
	
\bibitem{Lozano:2019jza}
  Y.~Lozano, N.~T.~Macpherson, C.~Nunez and A.~Ramirez,
  ``1/4 BPS AdS$_3$/CFT$_2$,''
  arXiv:1909.09636 [hep-th].
	
\bibitem{Lozano:2019zvg}
  Y.~Lozano, N.~T.~Macpherson, C.~Nunez and A.~Ramirez,
  ``Two dimensional ${\cal N}=(0,4)$ quivers dual to AdS$_3$ solutions in massive IIA,''
  arXiv:1909.10510 [hep-th].
	
\bibitem{Lozano:2019ywa}
  Y.~Lozano, N.~T.~Macpherson, C.~Nunez and A.~Ramirez,
  ``AdS$_3$ solutions in massive IIA, defect CFTs and T-duality,''
  JHEP {\bf 1912} (2019) 013
  doi:10.1007/JHEP12(2019)013
  [arXiv:1909.11669 [hep-th]].
	
\bibitem{Beck:2017wpm}
  S.~Beck, U.~Gran, J.~Gutowski and G.~Papadopoulos,
  ``All Killing Superalgebras for Warped AdS Backgrounds,''
  JHEP {\bf 1812} (2018) 047
  doi:10.1007/JHEP12(2018)047
  [arXiv:1710.03713 [hep-th]].
	
\bibitem{Deger:2019tem}
  N.~S.~Deger, C.~Eloy and H.~Samtleben,
  ``${\mathcal{N}=(8,0)}$ AdS vacua of three-dimensional supergravity,''
  JHEP {\bf 1910} (2019) 145
  doi:10.1007/JHEP10(2019)145
  [arXiv:1907.12764 [hep-th]].
	
		\bibitem{Dibitetto:2018ftj}
  G.~Dibitetto, G.~Lo Monaco, A.~Passias, N.~Petri and A.~Tomasiello,
  ``AdS$_3$ Solutions with Exceptional Supersymmetry,''
  Fortsch.\ Phys.\  {\bf 66} (2018) no.10,  1800060
  doi:10.1002/prop.201800060
  [arXiv:1807.06602 [hep-th]].
	
\bibitem{DeLuca:2018buk}
  G.~B.~De Luca, G.~L.~Monaco, N.~T.~Macpherson, A.~Tomasiello and O.~Varela,
  ``The geometry of $ \mathcal{N}=3 $ AdS$_{4}$ in massive IIA,''
  JHEP {\bf 1808} (2018) 133
  doi:10.1007/JHEP08(2018)133
  [arXiv:1805.04823 [hep-th]].
	
\bibitem{Dibitetto:2017klx}
  G.~Dibitetto and N.~Petri,
 ``6d surface defects from massive type IIA,''
  JHEP {\bf 1801} (2018) 039
  doi:10.1007/JHEP01(2018)039
  [arXiv:1707.06154 [hep-th]].

\bibitem{Macpherson:2018mif}
  N.~T.~Macpherson,
 ``Type II solutions on AdS$_{3} \times$ S$^{3} \times$ S$^{3}$ with large superconformal symmetry,''
  JHEP {\bf 1905} (2019) 089
  doi:10.1007/JHEP05(2019)089
  [arXiv:1812.10172 [hep-th]].
	
	
\bibitem{Macpherson:2017mvu}
  N.~T.~Macpherson, J.~Montero and D.~Prins,
  ``Mink $_3\times S^3$ solutions of type II supergravity,''
  Nucl.\ Phys.\ B {\bf 933} (2018) 185
  doi:10.1016/j.nuclphysb.2018.05.021
  [arXiv:1712.00851 [hep-th]].
	
\bibitem{Eberhardt:2018sce}
  L.~Eberhardt and I.~G.~Zadeh,
  ``$\mathcal{N}=(3,3)$ holography on ${\rm AdS}_3 \times ({\rm S}^3 \times {\rm S}^3 \times {\rm S}^1)/\mathbb Z_2$,''
  JHEP {\bf 1807} (2018) 143
  doi:10.1007/JHEP07(2018)143
  [arXiv:1805.09832 [hep-th]].
	
\bibitem{Prins:2013wza}
  D.~Prins and D.~Tsimpis,
  ``Type IIA supergravity and M -theory on manifolds with SU(4) structure,''
  Phys.\ Rev.\ D {\bf 89} (2014) 064030
  doi:10.1103/PhysRevD.89.064030
  [arXiv:1312.1692 [hep-th]].


\end{thebibliography}
\end{document}